\documentclass[twocolumn]{IEEEtran}
\usepackage{amsbsy}
\usepackage{latexsym}
\usepackage{amsmath}
\usepackage{ams}
\usepackage{times}
\input epsf
\begin{document}
\title{Adaptive Interference Suppression for CDMA Systems using Interpolated FIR Filters
with Adaptive Interpolators in Multipath Channels}

\author{Rodrigo C. de Lamare and Raimundo Sampaio-Neto  \\
\thanks{\footnotesize This work was supported by the Brazilian Council for Scientific and
Technological Development (CNPq). Dr. R. C. de Lamare was with
CETUC/PUC-RIO and is now a Lecturer with the Communications
Research Group, Department of Electronics, University of York,
York Y010 5DD, United Kingdom and Prof. R. Sampaio-Neto is with
CETUC/PUC-RIO, 22453-900, Rio de Janeiro, Brazil. Phone:
+55-21-31141701 Fax: +55-21-22945748. E-mails:
rcdl500@ohm.york.ac.uk, raimundo@cetuc.puc-rio.br} }

\maketitle
\thispagestyle{empty}
\begin{abstract}
In this work we propose an adaptive linear receiver structure
based on interpolated finite impulse response (FIR) filters with
adaptive interpolators for direct sequence code division multiple
access (DS-CDMA) systems in multipath channels. The interpolated
minimum mean-squared error (MMSE) and the interpolated constrained
minimum variance (CMV) solutions are described for a novel scheme
where the interpolator is rendered time-varying in order to
mitigate multiple access interference (MAI) and multiple-path
propagation effects. Based upon the interpolated MMSE and CMV
solutions we present computationally efficient stochastic gradient
(SG) and exponentially weighted recursive least squares type (RLS)
algorithms for both receiver and interpolator filters in the
supervised and blind modes of operation. A convergence analysis of
the algorithms and a discussion of the convergence properties of
the method are carried out for both modes of operation. Simulation
experiments for a downlink scenario show that the proposed
structures achieve a superior BER convergence and steady-state
performance to previously reported reduced-rank receivers at lower
complexity.


\end{abstract}

\section{Introduction}

\PARstart{A}{daptive} linear receivers
\cite{1,miller,madhow,rapajic} are a highly effective structure
for combatting interference in DS-CDMA systems, since they usually
show good performance and have simple adaptive implementation. The
linear minimum mean-squared error (MMSE) receiver
\cite{madhow,rapajic} implemented with an adaptive filter is one
of the most prominent design criteria for DS-CDMA systems. Such
receiver only requires the timing of the desired user and a
training sequence in order to suppress interference. Conversely,
when a receiver loses track of the desired user and a training
sequence is not available, a blind linear minimum variance (MV)
receiver \cite{honig,xu&tsatsanis} that trades off the need for a
training sequence in favor of the knowledge of the desired user's
spreading code can be used to retrieve the desired signal.

The works in \cite{1}-\cite{xu&tsatsanis} were restricted to
systems with short codes, where the spreading sequences are
periodic. However, adaptive techniques are also applicable to
systems with long codes provided some modifications are carried
out. The designer can resort to chip equalization \cite{klein}
followed by a despreader for downlink scenarios. For an uplink
solution, channel estimation algorithms for aperiodic sequences
\cite{buzzi,xu-long} are required and the sample average approach
for estimating the covariance matrix ${\bf R}=E[{\bf r}(i){\bf
r}^{H}(i)]$ of the observed data ${\bf r}(i)$ has to be replaced
by $\hat{\bf R} = {\bf P}{\bf P}^{H} + \sigma^2{\bf I}$, which is
constructed with a matrix ${\bf P}$ containing the effective
signature sequence of users and the variance $\sigma^{2}$ of the
receiver noise \cite{liu&xu}. In addition, with some recent
advances in random matrix theory \cite{Li} it is also possible to
deploy techniques originally developed for systems with short
codes in implementations with long codes. Furthermore, the
adaptive receiver structures reported so far
\cite{1}-\cite{liu&xu} can be easily extended to asynchronous
systems in uplink connections. In presence of large relative
delays amongst the users, the observation window of each user
should be expanded in order to consider an increased number of
samples derived from the offsets amongst users. Alternatively for
small relative delays amongst users, the designer can utilize chip
oversampling to compensate for the random timing offsets. These
remedies imply in augmented filter lengths and consequently
increased computational complexity.

In this context, a problem arises when the processing gain used in
the system and the number of parameters for estimation is large.
In these scenarios, the receiver has to cope with difficulties
such as significant computational burden, increased amount of
training and poor convergence and tracking performance. In
general, when an adaptive filter with a large number of taps is
used to suppress interference, then it implies slow response to
changing interference and channel conditions. Reduced-rank
interference suppression for DS-CDMA
\cite{bar-ness}-\cite{goldstein} was originally motivated by
situations where the number of elements in the receiver is large
and it is desirable to work with fewer parameters for complexity
and convergence reasons. Early works in reduced-rank interference
suppression for DS-CDMA systems \cite{bar-ness,wang&poor,song&roy}
were based on principal components (PC) of the covariance matrix
${\bf R}$ of the observation data. This requires a computationally
expensive eigen-decomposition to extract the signal subspace which
leads to poor performance in systems with moderate to heavy loads.
An attempt to reduce the complexity of PC approaches was reported
in \cite{singh&milstein} with the partial despreading (PD) method,
where the authors report a simple technique that allows the
selection of the performance between the matched filter and the
full-rank MMSE receiver. A promising reduced-rank technique for
interference suppression, denoted multistage Wiener filter (MWF),
was developed by Goldstein {\it et al.} in \cite{gold&reed} and
was later extended to SG and recursive adaptive versions by Honig
and Goldstein in \cite{goldstein}. A problem with the MWF approach
is that, although less complex than the full-rank solution, it
still presents a considerable computational burden and numerical
problems for implementation. In this work, we present an
alternative reduced-rank interference suppression scheme based on
interpolated FIR filters with adaptive interpolators that gathers
simplicity, great flexibility, low complexity and high
performance.

The interpolated FIR (IFIR) filter is a single rate structure that
is mathematically related to signal decimation followed by
filtering with a reduced number of elements \cite{neuvo1},
\cite{neuvo2}. The basic idea is to exploit the coefficient
redundancy in order to remove a number of impulse response
samples, which are recreated using an interpolation scheme. The
savings are obtained by interpolating the impulse response and by
decimating the interpolated signal. This technique exhibits
desirable properties, such as guaranteed stability, absence of
limit cycles and low computational complexity. Thus, adaptive IFIR
(AIFIR) filters \cite{abou},\cite{resende} represent an
interesting alternative for substituting classical adaptive FIR
filters. In some applications they show better convergence rate
and can reduce the computational burden for filtering and
coefficient updating, due to the reduced number of adaptive
elements. These structures have been extensively applied in the
context of digital filtering, although their use for parameter
estimation in communications remains unexplored.

Interference suppression with IFIR filters and time-varying
interpolators with batch methods, which require matrix inversions,
were reported in \cite{delamare1}. In this work, we investigate
the suppression of MAI and intersymbol interference (ISI) with
adaptive IFIR filters (that do not need matrix inversions) for
both supervised and blind modes of operation in synchronous
DS-CDMA systems with short codes.  A novel AIFIR scheme where the
interpolator is rendered adaptive is discussed and designed with
both MMSE and MV design criteria. The new scheme, introduced in
\cite{delamare2,delamare3}, yields a performance superior to
conventional AIFIR schemes \cite{abou,resende} (where the
interpolator is fixed) and a faster convergence performance than
full-rank and other existing reduced-rank receivers.
Computationally efficient stochastic gradient (SG) and recursive
least squares (RLS) type adaptive algorithms are developed for the
new structure based upon the MMSE and MV performance criteria with
appropriate constraints to mitigate MAI, ISI and jointly estimate
the channel. The motivation for the novel structure is to exploit
the redundancy found in DS-CDMA signals that operate in multipath,
by removing a number of samples of the received signal and
retrieving them through interpolation. The gains in convergence
performance over full-rank solutions are obtained through the
reduction of the number of parameters for estimation, leading to a
faster acquisition of the required statistics of the method and a
smaller misadjustment noise provided by the smaller filter
\cite{1,miller}. Furthermore, the use of an adaptive interpolator
can provide a time-varying and rapid means of compensation for the
decimation process and the discarded samples. The novel scheme has
the potential and flexibility to consistently yield faster
convergence than the full-rank approach since the designer can
choose the number of adaptive elements during the transient
process and upon convergence increase the number of elements up to
the full-rank. Unlike PC techniques, our scheme is very simple
because it does not require eigen-decomposition and its
performance is not severely degraded when the system load is
increased. In contrast to PD, the adaptive AIFIR structure jointly
optimizes two filters, namely the interpolator and the
reduced-rank, resulting in reduced-rank filters with fewer taps
and faster convergence than PD since the interpolator helps with
the compensation of the discarded samples. In comparison with the
MWF the proposed scheme is simpler, more flexible and more
suitable for implementation because the MWF has numerical problems
in fixed point implementations.


A convergence analysis of the algorithms and a discussion of the
global convergence properties of the method, which are not treated
in \cite{delamare1}-\cite{delamare3}, are undertaken for both
modes of operation. Specifically, we study the convergence
properties of the proposed joint adaptive interpolator and
receiver scheme and conclude that it leads to an optimization
problem with multiple global minima and no local minima. In this
regard and based on the analyzed convergence properties of the
method, we show that the prediction of the excess mean square
error (MSE) of both blind and supervised adaptive algorithms is
rendered possible through the study of the MSE trajectory of only
one of the jointly optimized parameter vectors, i.e. the
interpolator or the reduced-rank filters. Then, using common
assumptions of the adaptive filtering literature, such as the
independence theory, we analyze the trajectory of the mean tap
vector of the joint optimization of the interpolator and the
receiver and MSE trajectory.  We also provide some mathematical
conditions which explain why the new scheme with SG and RLS type
algorithms is able to converge faster than the full-rank scheme.
Although the novel structure and algorithms are examined in a
synchronous downlink scenario with periodic signature sequences in
this work, it should be remarked that they can be extended to long
codes and asynchronous systems provided the designer adopts the
modifications explained in the works reported in
\cite{klein}-\cite{Li}.

This paper is organized as follows. Section II describes the
DS-CDMA system model. The linear interpolated receiver principle
and design criteria, namely the MMSE and constrained MV (CMV) are
described in Section III. Section IV is dedicated to the
derivation of adaptive algorithms and Section V is devoted to the
global convergence properties of the method and the convergence
analysis of the algorithms. Section VI presents and discusses the
simulation results and Section VII gives the concluding remarks.

\section{DS-CDMA system model}

Let us consider the downlink of a synchronous DS-CDMA system with
$K$ users, $N$ chips per symbol and $L_{p}$ propagation paths. The
signal broadcasted by the base station intended for user $k$ has a
baseband representation given by:
\begin{equation}
x_{k}(t)=A_{k}\sum_{i=-\infty}^{\infty}b_{k}(i)s_{k}(t-iT)
\end{equation}
where $b_{k}(i) \in \{\pm1\}$ denotes the $i$-th symbol for user
$k$, the real valued spreading waveform and the amplitude
associated with user $k$ are $s_{k}(t)$ and $A_{k}$, respectively.
The spreading waveforms are expressed by $s_{k}(t) =
\sum_{i=1}^{N}a_{k}(i)\phi(t-iT_{c})$, where $a_{k}(i)\in
\{\pm1/\sqrt{N} \}$, $\phi(t)$ is the chip waveform, $T_{c}$ is
the chip duration and $N=T/T_{c}$ is the processing gain. Assuming
that the receiver is synchronized with the main path, the
coherently demodulated composite received signal is
\begin{equation}
r(t)= \sum_{k=1}^{K}\sum_{l=0}^{L_{p}-1}
h_{l}(t)x_{k}(t-\tau_{l})+n(t)
\end{equation}
where $h_{l}(t)$ and $\tau_{l}$ are, respectively, the channel
coefficient and the delay associated with the $l$-th path.
Assuming that $\tau_{k,l} = lT_{c}$, the channel is constant
during each symbol interval and the spreading codes are repeated
from symbol to symbol, the received signal $ r(t)$ after filtering
by a chip-pulse matched filter and sampled at chip rate yields the
$M=N+L_{p}-1$ dimensional received vector
\begin{equation}
{\bf r}(i) = {\bf H}(i) \sum_{k=1}^{K}A_{k} {\bf S}_{k}{\bf
b}_{k}(i) + {\bf n}(i)
\end{equation}
where { ${\bf n}(i) = [n_{1}(i) ~\ldots~n_{M}i)]^{T}$} is the
complex Gaussian noise vector with { $E[{\bf n}(i){\bf n}^{H}(i)]
= \sigma^{2}{\bf I}$}, where $(\cdot)^{T}$ and $(\cdot)^{H}$
denotes transpose and Hermitian transpose, respectively, and {
$E[.]$} is the expected value, the $k$-th user symbol vector is
${\bf b}_{k}(i) = [b_{k}(i+L_{s}-1)~ \ldots ~ b_{k}(i)~ \ldots~
b_{k}(i-L_{s}+1)]^{T}$, where $L_{s}$ is the ISI span and the {
$((2L_{s}-1)\times N)\times (2L_{s}-1)$} matrix { ${\bf S}_{k}$}
with non-overlapping shifted versions of the signature of user $k$
is
\begin{equation}
{\bf S}_{k} = \left[\begin{array}{c c c c c c c}
{\bf s}_{k} & 0 & \ldots & 0  \\
 0 & {\bf s}_{k} & \ddots  &  0 \\
 \vdots & \vdots & \ddots & \vdots \\
 0 & \ldots & \ldots & {\bf s}_{k} \end{array}\right]
\end{equation}
where the signature sequence for the $k$-th user is { ${\bf s}_{k}
= [a_{k}(1)~ \ldots ~a_{k}(N)]^{T}$} and the { $  M~\times
((2L_{s}-1)\times N)$} channel matrix { ${\bf H}(i)$} is
\begin{equation}
{\bf H}(i) = \left[\hspace*{-0.85em}\begin{array}{c c c c c c}
h_{0}(i) &  \ldots & h_{L_{p}-1}(i) & \ldots & 0 & 0  \\
0 & h_{0}(i) &  \ldots & h_{L_{p}-1}(i) & \ldots & 0   \\
 \vdots  & \ddots & \ddots & \ddots & \ddots & \vdots \\
 0 & 0 &  \ldots & h_{0}(i) & \ldots & h_{L_{p}-1}(i) \hspace*{-0.85em} \end{array}\right]
\end{equation}
where $h_{l}(i) = h_{l}(iT_{c})$ for ~$l=0,\ldots,L_{p}-1$. The
MAI arises from the non-orthogonality between the received
signals, whereas the ISI span $L_{s}$ depends on the length of the
channel response, which is related to the length of the chip
sequence. For { $L_{p}=1,~ L_{s}=1$} (no ISI), for { $1<L_{p}\leq
N, L_{s}=2$}, for { $N <L_{p}\leq 2N, L_{s}=3$} and so on.

\section{Linear Interpolated CDMA receivers}

The underlying principles of the proposed CDMA receiver structure
are detailed here. Fig. 1 shows the structure of an IFIR receiver,
where an interpolator and a reduced-rank receiver that are
time-varying are employed. The $M\times 1$ received vector ${\bf
r}(i)=[r_{0}^{(i)}~ \ldots ~r_{M-1}^{(i)}]^{T}$ is filtered by the
interpolator filter ${\bf v}_{k}(i) = [v_{k,0}^{(i)} \ldots
v_{k,N_{I}-1}^{(i)}]^{T}$, yielding the interpolated received
vector ${\bf r}_{k}(i)$. The vector ${\bf r}_{k}(i)$ is then
projected onto an $M/L \times 1$ vector $\bar{\bf r}_{k}(i)$. This
procedure corresponds to removing $L-1$ samples of ${\bf
r}_{k}(i)$ of each set of $L$ consecutive ones. Then
 the inner product of $\bar{\bf r}_{k}(i)$ with the
$M/L$-dimensional vector of filter coefficients ${\bf
w}_{k}(i)=[w_{k,0}^{(i)}~\ldots~w_{k,M/L-1}^{(i)}]^{T}$ is
computed to yield the output $x_k(i)$.

\begin{figure}[!htb]
\begin{center}
\def\epsfsize#1#2{1\columnwidth}
\epsfbox{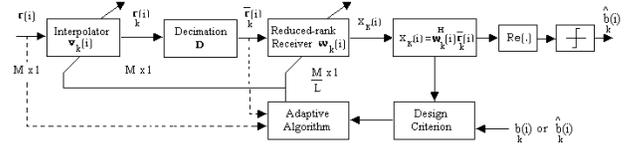} \caption{\small Proposed adaptive reduced-rank
receiver structure.}
\end{center}
\end{figure}

The projected interpolated observation vector $\bar{\bf r}_{k}(i)
= {\bf D} {\bf r}_{k}(i)$ is obtained with the aid of the
$M/L\times M$ projection matrix ${\bf D}$ which is mathematically
equivalent to signal decimation on the $M \times 1$ vector ${\bf
r}_{k}(i)$. An interpolated receiver with decimation factor $L$
can be designed by choosing {\bf D} as:
\begin{equation}
{\bf D} = \left[\hspace*{-1.25em}\begin{array}{c}
~~~1 ~~~~~  0 ~~~~ 0 ~~~~~~ 0 ~~~~~ 0 ~~~ \ldots ~~~~ 0 ~~~~ 0 ~~~~ 0 ~~~~ 0 ~~~~ 0 \\
~~~~\vdots ~~~~~~ \vdots ~~~~~ \vdots ~~~~~ \vdots ~~~~~~~ \vdots ~~~~~~~~ \vdots  ~~~~~~ \vdots ~~~~~ \vdots ~~~~~ \vdots ~~~~~ \vdots ~~~~~ \vdots \\
\underbrace{0~~ \ldots ~~ 0}_{(m-1)L~zeros}~~1 ~~~~~~ 0 ~~~ \ldots ~~~~ 0 ~~~~ 0 ~~~~ 0 ~~~~ 0 ~~~~ 0\\
~~~~~~\vdots ~~~~~~ \vdots ~~~~~ \vdots ~~~~~ \vdots ~~~~~~~ \vdots ~~~~ \ddots  ~~~~ \vdots ~~~~~ \vdots ~~~~~ \vdots ~~~~~ \vdots ~~~~~ \vdots~ \\
~~~~~~~\underbrace{0 ~~~~ 0 ~~~~ 0 ~~~~ 0 ~~~~~~~ 0 ~~~ \ldots ~~~~ 0}_{(M/L-1)L~zeros} ~~~ 1 ~~  \underbrace{0~~~  \ldots ~~  0}_{(L-1) ~zeros}~ \\
 \end{array}\right]
 \end{equation}
where $m$ ($m=1,2,\ldots, M/L$) denotes the $m$-th row. The
strategy, that allows us to devise solutions for both interpolator
and receiver, is to express the estimated symbol $x_{k}(i) = {\bf
w}^{H}_{k}(i)\bar{\bf r}_{k}(i)$ as a function of ${\bf w}_{k}(i)$
and ${\bf v}_{k}(i)$ (we will drop the subscript $k$ and symbol
index (i) for ease  of presentation):
\begin{equation}
\begin{split}
x_{k}(i) & = w_{0}^{*} {\bf v}^{H}_{k} \dot{\bf r}_{0} + w_{1}^{*}
{\bf v}^{H}_{k} \dot{\bf r}_{1}+ \ldots
 + w_{M/L-1}^{*} {\bf v}^{H}_{k}\dot{\bf r}_{M/L-1} \\
& = {\bf v}^{H}_{k}(i)\Big[~ \dot{\bf r}_{0}^{(i)}|~ \ldots ~
|~\dot{\bf r}_{M/L-1}^{(i)}~\Big]{\bf w}_{k}^{*}(i) = {\bf
v}^{H}_{k}(i)\boldsymbol{ \Re}(i){\bf w}^{*}_{k}(i)
\end{split}
\end{equation}
where ${\bf u}_{k}(i)=\boldsymbol{ \Re}(i){\bf w}^{*}_{k}(i)$ is
an $N_{I}\times 1$ vector, the $M/L$ coefficients of ${\bf
w}_{k}(i)$ and the $N_{I}$ elements of ${\bf v}_{k}(i)$ are
assumed to be complex, the asterisk denotes complex conjugation
and { $\dot{\bf r}_{s}(i)$} is a length $N_{I}$ segment of the
received vector { ${\bf r}(i)$} beginning at { ${ r}_{s\times
L}(i)$} and
\begin{equation}
\boldsymbol{ \Re}(i) = \left[\begin{array}{c c c c c}
r_{0}^{(i)} & r_{L}^{(i)}   & \ldots & r_{(M/L-1)L}^{(i)}  \\
r_{1}^{(i)}  & r_{L+1}^{(i)}   & \ldots & r_{(M/L-1)L+1}^{(i)}  \\
\vdots & \vdots  & \ddots & \vdots \\
r_{N_{I}-1}^{(i)}  & r_{L+N_{I}-1}^{(i)}  & \ldots & r_{(M/L-1)L+N_{I}-1}^{(i)}  \\
 \end{array}\right]
\end{equation}
The interpolated linear receiver design is equivalent to
determining an FIR filter ${\bf w}_{k}(i)$ with $M/L$ coefficients
that provide an estimate of the desired symbol:
\begin{equation}
 \hat{b}_{k}(i) = sgn\Big(Re\Big[{\bf w}_{k}^{H}(i)\bar{\bf r}_{k}(i)\Big]\Big)
\end{equation}
where $Re(\cdot)$ selects the real part, $sgn(\cdot)$ is the
signum function and the receiver parameter vector ${\bf w}_{k}$ is
optimized according to a selected design criterion.

\subsection{MMSE Reduced-Rank Interpolated Receiver Design}

The MMSE solutions for ${\bf w}_{k}(i)$ and ${\bf v}_{k}(i)$ can
be computed if we consider the optimization problem whose cost
function is
\begin{equation}
J_{MSE}({\bf w}_{k}(i),{\bf v}_{k}(i)) = E\Big[|b_{k}(i)- {\bf
v}^{H}_{k}(i) \boldsymbol{\Re}(i){\bf w}_{k}^{*}(i) |^{2}\Big]
\end{equation}
where $b_{k}(i)$ is the desired symbol for user $k$ at time index
$(i)$. By fixing the interpolator ${\bf v}_{k}(i)$ and minimizing
(10) with respect to ${\bf w}_{k}(i)$  the interpolated Wiener
filter/receiver weight vector is
\begin{equation}
{\bf w}_{k}(i) = \boldsymbol{\alpha} ({\bf v}_{k})=\bar{\bf
R}^{-1}_{k}(i) \bar{\bf p}_{k}(i)
\end{equation}
where $\bar{\bf R}_{k}(i) = E[\bar{\bf r}_{k}(i) \bar{\bf
r}_{k}^{H}(i)]$, $\bar{\bf p}_{k}(i) = E[b_{k}^{*}(i)\bar{\bf
r}_{k}(i)]$, $\bar{\bf r}_{k}(i) = \boldsymbol{ \Re}^{T}(i){\bf
v}_{k}^{*}(i)$ and by fixing ${\bf w}_{k}(i)$ and minimizing (10)
with respect to ${\bf v}_{k}(i)$ the interpolator weight vector is
\begin{equation} {\bf v}_{k}(i) = \boldsymbol{\beta} ({\bf w}_{k}) = {\bf
R}^{-1}_{u_{k}}(i) {\bf p}_{u_{k}}(i)
\end{equation}
where  ${\bf R}_{u_{k}}(i) = E[{\bf u}_{k}(i){\bf u}^{H}_{k}(i)]$,
${\bf p}_{u_{k}}(i) = E[b_{k}^{*}(i){\bf u}_{k}(i)]$ and ${\bf
u}_{k}(i) = \boldsymbol{\Re}(i){\bf w}_{k}^{*}(i)$. The associated
MSE expressions are
\begin{equation}
J({\bf v}_{k}) = J_{MSE}(\boldsymbol{\alpha}({\bf v}_{k}),{\bf
v}_{k}) = \sigma^{2}_{b} - \bar{\bf p}^{H}_{k}(i) {\bf
R}^{-1}_{k}(i) \bar{\bf p}_{k}(i)
\end{equation}
\begin{equation}
J_{MSE}({\bf w}_{k},\boldsymbol{\beta}({\bf w}_{k})) =
\sigma^{2}_{b} - {\bf p}^{H}_{u_{k}}(i){\bf R}^{-1}_{u_{k}}(i)
{\bf p}_{u_{k}}(i)
\end{equation}
where $\sigma^{2}_{b}=E[|b(i)|^{2}]$. Note that points of global
minimum of (10) can be obtained by ${\bf v}_{k,opt}= \arg
\min_{{\bf v}_{k}}~ J({\bf v}_{k})$ and ${\bf
w}_{k,opt}=\boldsymbol{\alpha}({\bf v}_{k,opt})$ or ${\bf
w}_{k,opt}= \arg \min_{{\bf w}_{k}} J_{MSE}({\bf
w}_{k},\boldsymbol{\beta}({\bf w}_{k}))$ and ${\bf
v}_{k,opt}=\boldsymbol{\beta}({\bf w}_{k,opt})$. At the minimum
point (13) equals (14) and the MMSE for the proposed structure is
achieved. We remark that (11) and (12) are not closed-form
solutions for ${\bf w}_{k}(i)$ and ${\bf v}_{k}(i)$ since (11) is
a function of ${\bf v}_{k}(i)$ and (12) depends on ${\bf
w}_{k}(i)$ and thus it is necessary to iterate (11) and (12) with
an initial guess to obtain a solution, as in \cite{delamare1}. An
iterative MMSE solution can be sought via adaptive algorithms.

\subsection{Constrained Minimum Variance (CMV) Reduced-Rank Interpolated Receiver Design}

The interpolated CMV receiver parameter vector ${\bf w}_{k}$ and
the interpolator parameter vector ${\bf v}_{k}$ are obtained by
minimizing
\begin{equation}
\begin{split}
J_{MV}({\bf w}_{k},{\bf v}_{k}) & = E\Big[|x_{k}(i)|^{2}\Big]
=E\Big[|{\bf v}_{k}^{H}(i) \boldsymbol{ \Re}(i){\bf
w}_{k}^{*}(i)|^{2}\Big] \\ & = {\bf w}_{k}^{H}(i)\bar{\bf
R}_{k}{\bf w}_{k}(i) = {\bf v}_{k}^{H}(i){\bf R}_{\bf u_{k}}{\bf
v}_{k}(i)
\end{split}
\end{equation}
subject to the proposed constraints { ${\bf C}_{k}^{H}{\bf D}^{H}
{\bf w}_{k}(i) = {\bf g}(i)$} and {$||{\bf v}_{k}(i)||=1$}, where
the $M \times L_{p}$ constraint matrix ${\bf C}_{k}$
contains one-chip shifted versions of the signature sequence of
user $k$, ${\bf g}(i)$ is an $L_{p}$-dimensional parameter vector
to be determined. The vector of constraints { ${\bf g}(i)$} can be
chosen amongst various criteria although in this work we adopt {
${\bf g}(i)$} as the channel parameter vector (${\bf g}= [h_{0}
\ldots h_{L_{p}-1}]^{T}$) because it provides better performance
than other choices as reported in \cite{xu&tsatsanis}. The
proposed constraint { $||{\bf v}_{k}(i)||=1$} ensures adequate
design values for the interpolator filter ${\bf v}_{k}$, whereas {
${\bf C}_{k}^{H}{\bf D}^{H}{\bf w}_{k}(i) = {\bf g}(i)$} avoids
the suppression of the desired signal. By fixing ${\bf v}_{k}$,
taking the gradient of the Lagrangian function $J_{MV}^{l}({\bf
w}_{k},{\bf v}_{k}) = E\Big[ [|{\bf v}_{k}^{H}(i) \boldsymbol{
\Re}(i){\bf w}_{k}^{*}(i)|^{2}\Big] +Re\Big[ ({\bf C}_{k}^{H}{\bf
D}^{H}{\bf w}_{k}(i)-{\bf g}(i) )^{H}\boldsymbol{\lambda} \Big]$,
where $\boldsymbol{\lambda}$ is a vector of Lagrange multipliers,
with respect to ${\bf w}_{k}$ and setting it to ${\bf 0}$ we get:
$$ E\Big[  \bar{\bf r}_{k}(i)\bar{\bf r}_{k}^{H}(i)\Big] {\bf w}_{k}(i) + {\bf D C}_{k} \boldsymbol{\lambda} = {\bf 0} $$
$$\Longrightarrow ~ {\bf w}_{k}(i) = -\bar{\bf R}_{k}^{-1}(i) {\bf
D C}_{k} \boldsymbol{\lambda}$$ Using the constraint ${\bf
C}_{k}^{H}{\bf D}^{H}{\bf w}_{k}(i) = {\bf g}(i)$ and substituting
${\bf w}_{k}(i) = -\bar{\bf R}_{k}^{-1}(i) {\bf D C}_{k}
\boldsymbol{\lambda}$ we arrive at \\ $\boldsymbol{\lambda} =
-({\bf C}_{k}^{H}{\bf D}^{H}\bar{\bf R}_{k}^{-1}(i){\bf D
C}_{k})^{-1} {\bf g}_{k}(i)$. The resulting expression for the
receiver is
\begin{equation}
{\bf w}_{k}(i) = \boldsymbol{\alpha}_{o}({\bf v}_{k})=\bar{\bf
R}_{k}(i)^{-1} {\bf DC}_{k}({\bf C}_{k}^{H}{\bf D}^{H}\bar{\bf
R}_{k}(i)^{-1} {\bf DC}_{k})^{-1}{\bf g}(i)
\end{equation}
and the associate minimum output variance is
\begin{equation}
\begin{split}
J_{o}({\bf v}_{k}) &  = J_{MV}(\boldsymbol{\alpha}_{o}({\bf
v}_{k}),{\bf v}_{k})  = {\bf w}_{k}^{H}(i)\bar{\bf R}_{k}(i){\bf
w}_{k}(i) \\ & = {\bf g}^{H}(i)({\bf C}^{H}_{k}{\bf D}^{H}\bar{\bf
R}_{k}(i)^{-1} {\bf DC}_{k})^{-1}{\bf g}(i)
\end{split}
\end{equation}
By fixing ${\bf w}_{k}$, the solution that minimizes (15) is:
\begin{equation}
{\bf v}_{k}(i) = \boldsymbol{\beta}_{o}({\bf w}_{k})= \arg
\min_{{\bf v}}~~ {\bf v}^{H}{\bf R}_{{\bf u}_{k}}(i){\bf v}
\end{equation}
subject to $||{\bf v}_{k}(i)||=1$. Therefore, the solution for the
interpolator is the normalized eigenvector of ${\bf R}_{\bf
u_{k}}$ corresponding to its minimum eigenvalue, via singular
value decomposition (SVD). As occurs with the MMSE approach we
iterate (16) and (18) with an initial guess to obtain a CMV
solution \cite{delamare1}. Note also that (16) assumes the
knowledge of the channel parameters. However, in applications
where multipath is present these parameters are not known and thus
channel estimation is required. To blindly estimate the channel we
use the method of \cite{xu&tsatsanis,power}:
\begin{equation}
\hat{\bf g}(i) = \arg \min_{\bf g}~~{{\bf g}^{H} {\bf
C}^{H}_{k}{\bf R}^{-m}(i){\bf C}_{k}{\bf g}}
\end{equation}
subject to $||\hat{\bf g}||=1$, where ${\bf R}(i)=E[{\bf r}(i){\bf
r}^{H}(i)]$, $m$ is a finite power and whose solution is the
eigenvector corresponding to the minimum eigenvalue of the
$L_{p}\times L_{p}$ matrix { $ {\bf C}^{T}_{k}{\bf R}(i)^{-m}{\bf
C}_{k}$} through SVD. Note that in this work we restrict the
values of $m$ to $1$ although the performance of the channel
estimator and consequently of the receiver can be improved by
increasing $m$. In the next section, we propose iterative
solutions via adaptive algorithms.

\section{Adaptive algorithms}

We describe SG and RLS algorithms \cite{haykin} (see Chapters $9$
and $13$) that adjust the parameters of the receiver and the
interpolator based on the MMSE criterion and the constrained
minimization of the MV cost function \cite{delamare2,delamare3}.
The novel structure, shown in Fig. 1 and denoted INT, for the
receivers gathers fast convergence, low complexity and additional
flexibility since the designer can adjust the decimation factor
$L$ and the length of the interpolator $N_{I}$ depending on the
needs of the application and the hostility of the environment.
Based upon the MMSE and CMV design criteria, the proposed receiver
structure has the following modes of operation: training mode,
where it employs a known training sequence; decision directed
mode, which uses past decisions in order to estimate the receiver
parameters; and blind mode, which is based on the CMV criterion
and trades-off the training sequence against the knowledge of the
signature sequence. The complexity in terms of arithmetic
operations of the algorithms associated with the INT and the
existing techniques is included as a function of the number of
adaptive elements for comparison purposes.

\subsection{Least mean squares (LMS) algorithm  }

Given the projected interpolated observation vector { $\bar{\bf
r}_{k}(i)$} and the desired symbol $b_{k}(i)$, we consider the
following cost function:
\begin{equation}
J_{MSE} = |b_{k}(i)- {\bf v}^{H}_{k}(i) \boldsymbol{\Re}(i){\bf
w}_{k}^{*}(i) |^{2}
\end{equation}
Taking the gradient terms of (20) with respect to ${\bf
w}_{k}(i)$, ${\bf v}_{k}(i)$ and using the gradient descent rules
\cite{haykin} (see Chapter $9$, pp. 367-371) ${\bf w}_{k}(i+1) =
{\bf w}_{k}(i) - \mu \nabla^{J_{MSE}}_{\bf w^{*}}$ and ${\bf
v}_{k}(i+1) = {\bf v}_{k}(i) - \eta \nabla^{J_{MSE}}_{\bf v^{*}}$
yields:
\begin{equation}
{\bf v}_{k}(i+1) = {\bf v}_{k}(i) + \eta e_{k}^{*}(i){\bf
u}_{k}(i)
\end{equation}
\begin{equation}
{\bf w}_{k}(i+1) = {\bf w}_{k}(i) + \mu e_{k}^{*}(i)\bar{\bf
r}_{k}(i)
\end{equation}
where $e_{k}(i)=b_{k}(i) - {\bf w}_{k}(i)^{H}\bar{\bf r}_{k}(i)$
is the error for user $k$, {${\bf u}_{k} = \boldsymbol{
\Re}(i){\bf w}_{k}(i)$}, $\mu$ and $\eta$ are the step sizes of
the algorithm for ${\bf w}_{k}(i)$ and ${\bf v}_{k}(i)$. The LMS
algorithm for the proposed structure described in this section has
a computational complexity $O(M/L + N_{I})$. In fact, the proposed
structure trades off one LMS algorithm with complexity $O(M)$
against two LMS algorithms with complexity $O(M/L)$ and
$O(N_{I})$, operating in parallel. It is worth noting that, for
stability and to facilitate tuning of parameters, it is useful to
employ normalized step sizes and consequently NLMS type recursions
when operating in a changing environment and thus we have $\mu(i)
= \frac{\eta_{0}}{ \bar{\bf r}_{k}^{H}(i)\bar{\bf r}_{k}(i)}$ and
$\eta(i) = \frac{\mu_{0}}{{\bf u}_{k}^{H}(i){\bf u}_{k}(i)}$ as
the step sizes of the algorithm for ${\bf w}_{k}(i)$ and ${\bf
v}_{k}(i)$, where ${\mu_{0}}$ and ${\eta_{0}}$ are the convergence
factors.

\subsection{Recursive least squares (RLS) algorithm }

Consider the time average estimate of the matrix ${\bar{\bf
R}}_{k}$, required in (11), given by ${\hat{\bar{\bf R}}}_{k}(i) =
\sum_{l=1}^{i}\alpha^{i-l}{\bar{\bf r}}_{k}(l){\bar{\bf
r}}_{k}^{H}(l)$, where $\alpha$ ($0<\alpha \leq 1$) is the
forgetting factor, that can be alternatively expressed by {
${\hat{\bar{\bf R}}}_{k}(i) = \alpha{\hat{\bar{\bf R}}}_{k}(i-1) +
{\bar{\bf r}}_{k}(i){\bar{\bf r}}_{k}^{H}(i)$}. To avoid the
inversion of ${\hat{\bar{\bf R}}}_{k}(i)$ required in (11), we use
the matrix inversion lemma and define  ${\bf P}_{k}(i) =
{\hat{\bar{ \bf R}}}^{-1}_{k}(i)$ and the gain vector ${\bf
G}_{k}(i)$ as:
\begin{equation}
{\bf G}_{k}(i) =  \frac{\alpha^{-1}{\bf P}_{k}(i-1)\bar{\bf
r}_{k}(i)} {1+ \alpha^{-1} \bar{\bf r}_{k}^{H}(i)  {\bf
P}_{k}(i-1) \bar{\bf r}_{k}(i)}
\end{equation}
and thus we can rewrite { ${\bf P}_{k}(i)$} as
\begin{equation}
{\bf P}_{k}(i) = \alpha^{-1}{\bf P}_{k}(i-1)-\alpha^{-1}{\bf
G}_{k}(i) \bar{\bf r}^{H}_{k}(i) {\bf P}_{k}(i-1)
\end{equation}
By rearranging (23) we have {${\bf G}_{k}(i) = \alpha^{-1}{\bf
P}_{k}(i-1)\bar{\bf r}_{k}(i)-\alpha^{-1}{\bf G}_{k}(i) \bar{\bf
r}_{k}^{H}(i) {\bf P}_{k}(i-1)\bar{\bf r}_{k}(i)= {\bf
P}_{k}(i)\bar{\bf r}_{k}(i)$ }. By employing the LS solution (a
time average of (11)) and the recursion $\hat{\bf p}_{k}(i) =
\alpha \hat{\bf p}_{k}(i-1) + \bar{\bf r}_{k}(i)b_{k}^{*}(i)$ we
obtain
\begin{equation}
{\bf w}_{k}(i) = {\hat{\bar{\bf R}}}^{-1}_{k}(i){\hat{\bf
p}}_{k}(i) = \alpha{\bf P}_{k}(i){\hat{\bf p}}_{k}(i-1)+{\bf
P}_{k}(i)\bar{\bf r}_{k}(i)b_{k}^{*}(i)
\end{equation}
Substituting (24) into (25) yields:
\begin{equation}
{\bf w}_{k}(i) = {\bf w}_{k}(i-1)+{\bf G}_{k}(i)\xi_{k}^{*}(i)
\end{equation}
where the {\it a priori} estimation error is described by
$\xi_{k}(i) = b_{k}(i) - {\bf w}_{k}^{H}(i-1)\bar{\bf r}_{k}(i)$.
Similar recursions for the interpolator are devised by using (12).
The estimate ${\hat{{\bf R}}}_{{\bf u}_{k}}$ can be obtained
through ${\hat{\bf R}}_{{\bf u}_{k}}(i)= \sum_{l=1}^{i}
\alpha^{i-l} {\bf u}_{k}(l){\bf u}_{k}^{H}(l)$ and can be
alternatively written as { ${\hat{\bf R}}_{{\bf u}_{k}}(i) =
\alpha {\hat{\bf R}}_{{\bf u}_{k}}(i-1) + {\bf u}_{k}(i){\bf
u}_{k}^{H}(i)$}. To avoid the inversion of ${\hat{\bf R}}_{{\bf
u}_{k}}$ we use the matrix inversion lemma and again for
convenience of computation we define ${\bf P}_{{\bf
u}_{k}}(i)={\hat{\bf R}}_{{\bf u}_{k}}^{-1}(i)$ and the Kalman
gain vector ${\bf G}_{{\bf u}_{k}}(i)$ as:
\begin{equation}
{\bf G}_{{\bf u}_{k}}(i) =  \frac{\alpha^{-1}{\bf P}_{{\bf
u}_{k}}(i-1){\bf u}_{k}(i)} {1+ \alpha^{-1} {\bf u}_{k}^{H}(i){\bf
P}_{{\bf u}_{k}}(i-1) {\bf u}_{k}(i)}
\end{equation}
and thus we can rewrite (27) as
\begin{equation}
{\bf P}_{{\bf u}_{k}}(i) = \alpha^{-1}{\bf P}_{{\bf
u}_{k}}(i-1)-\alpha^{-1}{\bf G}_{{\bf u}_{k}}(i) {\bf
u}^{H}_{k}(i){\bf P}_{{\bf u}_{k}}(i-1)
\end{equation}
By proceeding in a similar approach to the one taken to obtain
(26) we arrive at
\begin{equation}
{\bf v}_{k}(i) = {\bf v}_{k}(i-1)+{\bf G}_{{\bf
u}_{k}}(i)\xi_{k}^{*}(i)
\end{equation}
The RLS algorithm for the proposed structure trades off a
computational complexity of $O(M^{2})$ against two RLS algorithms
operating in parallel, with complexity $O((M/L)^{2})$ and
$O(N_{I}^{2})$, respectively. Because $N_{I}$ is small
($N_{I}<<M$, as will be shown later) the computational advantage
of the RLS combined with the INT structure is rather significant.

\subsection{Constrained minimum variance stochastic gradient (CMV-SG) algorithm }
Consider the unconstrained Lagrangian MV cost function:
\begin{equation}
\begin{split}
J_{MV} & = ({\bf v}_{k}^{H}(i){\bf u}_{k}(i){\bf u}_{k}^{H}(i){\bf
v}_{k}(i)) +\boldsymbol{\lambda}^{H} ( {\bf C}_{k}^{H} {\bf
D}^{H}{\bf w}_{k}(i)  -{\bf g}(i)) \\ & \quad +  (  {\bf
w}_{k}^{H}(i){\bf D}{\bf C}_{k} -{\bf
g}^{H}(i))\boldsymbol{\lambda}
\end{split}
\end{equation}
where $\boldsymbol{ \lambda}$ is a vector of Lagrange multipliers.
An SG solution can be devised by taking the gradient terms of (30)
with respect to ${\bf w}_{k}(i)$ and ${\bf v}_{k}(i)$ as described
by { ${\bf w}_{k}(i+1) = {\bf w}_{k}(i) - \mu_{w}(i)\nabla J_{{\bf
w}_{k}(i)}$, ${\bf v}_{k}(i+1) = {\bf v}_{k}(i) - \eta(i)\nabla
J_{{\bf v}_{k}(i)}$} which adaptively minimizes $J_{MV}$ with
respect to ${\bf w}_{k}(i)$ and ${\bf v}_{k}(i)$. Substituting the
gradient terms the equations become
\begin{equation}
{\bf w}_{k}(i+1) = {\bf w}_{k}(i) - \mu_{w}(i)(
x_{k}^{*}(i)\bar{\bf r}_{k}(i) +{\bf DC}_{k}
\boldsymbol{\lambda}(i)) \end{equation}
\begin{equation}
{\bf v}_{k}(i+1) = {\bf v}_{k}(i) - \eta(i) x_{k}^{*}(i) {\bf
u}_{k}(i)
\end{equation}
where $x_{k}(i) = {\bf w}_{k}^{H}(i)\bar{\bf r}_{k}(i)={\bf
v}_{k}^{H}(i){\bf u}_{k}(i)$. We use (32) and can make {  ${\bf
v}_{k}(i+1)\leftarrow {\bf v}_{k}(i+1)/||{\bf v}_{k}(i+1)||$} to
update the interpolator ${\bf v}_{k}$. It is worth noting that in
our studies the normalization on SG algorithms does not lead to
different results from the ones obtained with a non-normalized
interpolator recursion. In this regard, analyzing the convergence
of (32) without normalization is mathematically simpler and gives
us the necessary insight into its convergence. By combining the
constraint ${\bf C}_{k}{\bf D}^{H}{\bf w}_{k}(i) = {\bf g}(i)$ and
(32) we obtain the Lagrange multiplier
\begin{equation}
\boldsymbol{\lambda}(i) = ({\bf C}^{H}_{k}{\bf D}^{H}{\bf D}{\bf
C}_{k})^{-1}\times ({\bf C}^{H}{\bf Dw}_{k}(i) - \mu_{w}{\bf
C}^{H}{\bf D}x_{k}^{*}(i) \bar{\bf r}_{k}(i)-{\bf
g}(i))\end{equation} By substituting (33) into (31) we arrive at
the update rules for the estimation of the parameters of the
receiver ${\bf w}_{k}$:
\begin{equation} {\bf w}_{k}(i+1) = \boldsymbol{\Pi}_{k}({\bf w}_{k}(i) - \mu_{w}(i)x_{k}^{*}(i)\bar{\bf r}_{k}(i))+
{\bf DC}_{k} ({\bf C}_{k}^{H}{\bf D}^{H}{\bf D}{\bf
C}_{k})^{-1}{\bf g}(i)
\end{equation}
where $\boldsymbol{\Pi}_{k} = {\bf I} -{\bf DC}_{k}({\bf
C}_{k}^{H} {\bf D}^{H}{\bf D}{\bf C}_{k})^{-1}{\bf C}_{k}^{H}{\bf
D}^{H}$ is a matrix that projects ${\bf w}_{k}$ onto another
hyperplane to ensure the constraints.

Normalized versions of these algorithms can be devised by
substituting (32) and (34) into the MV cost function,
differentiating the cost function with respect to {  $\mu_{w}(i)$}
and {  $\mu_{v}(i)$}, setting them to zero and solving the new
equations. Hence, the CMV-SG algorithm proposed here for the INT
receiver structure adopts the normalized step sizes {  $\mu_{w}(i)
= \frac{\mu_{0}}{\bar{\bf
r}_{k}^{H}(i)\boldsymbol{\Pi}_{k}\bar{\bf r}_{k}(i)}$} and {
$\eta(i) = \frac{\eta_{0}}{{\bf u}_{k}^{H}{\bf u}_{k}(i)}$} where
{ $\mu_{0}$} and {  $\eta_{0}$} are the convergence factors for {
${\bf w}_{k}$} and {  ${\bf v}_{k}$}, respectively.

The channel estimate $\hat{\bf g}(i)$ is obtained through the
power method and the SG technique described in \cite{power}. The
method is an SG adaptive version of the blind channel estimation
algorithm described in (19) and introduced in \cite{power2} that
requires only $O(L_{p}^2)$ arithmetic operations to estimate the
channel, against $O(L_{p}^{3})$ of its SVD version. In terms of
computational complexity, for the rejection of MAI and ISI, the
proposed blind interpolated receiver trades off one blind
algorithm with complexity {  $O(M)$} against two blind algorithms
with complexity {  $O(M/L)$} and { $O(N_{I})$}, operating in
parallel.

\subsection{Constrained minimum variance recursive least squares (CMV-RLS) algorithm}

Based upon the expressions for the receiver ${\bf w}_{k}$ and
interpolator ${\bf v}_{k}$ in (16) and (18) of the interpolated
CMV receiver, we develop a computationally efficient RLS algorithm
for the INT structure that estimates the parameters of ${\bf
w}_{k}$ and ${\bf v}_{k}$.

The iterative power method \cite{golub}(see pp.
$405-408$),\cite{watkins}(see pp. $314-333$) is used in numerical
analysis to compute the eigenvector corresponding to the largest
singular value of a matrix. In order to obtain an estimate of
${\bf v}_{k}$ and avoid the SVD on the estimate of the matrix
${\bf R}_{{\bf u}_{k}}(i)$ we resort to a variation of the
iterative power method to obtain the eigenvector of ${\bf R}_{{\bf
u}_{k}}(i)$ which corresponds to the minimum eigenvalue.

Specifically, we apply the power method to the difference between
${\bf R}_{{\bf u}_{k}}(i)$ and the identity matrix ${\bf I}$,
rather than applying it to the inverse of ${\bf R}_{{\bf
u}_{k}}(i)$. This approach, known as shift iterations
\cite{watkins}(see pp. $319$), leads to computational savings on
one order of magnitude since direct SVD requires {
$O(N_{I}^{3})$}, while our approach needs { $O(N_{I}^{2})$}. The
simulations carried out reveal that this method exhibits no
performance loss. Hence, we estimate ${\bf R}_{{\bf u}_{k}}(i)$
via the recursion $\hat{\bf R}_{{\bf u}_{k}}(i) =\sum_{n=0}^{i}
\alpha^{i-n}{\bf u}_{k}(n){\bf u}^{H}_{k}(n)$ and then obtain the
interpolator $\hat{\bf v}_{k}$ with a one step iteration given by:
\begin{equation}
\hat{\bf v}_{k}(i)= ({\bf I}-\nu_{k}(i)\hat{\bf R}_{{\bf
u}_{k}}(i))\hat{\bf v}_{k}(i-1)
\end{equation}
where $\nu_{k}(i)=1/tr[\hat{\bf R}_{{\bf u}_{k}}(i)]$. After that,
we make $\hat{\bf v}_{k}(i)\leftarrow \hat{\bf
v}_{k}(i)/||\hat{\bf v}_{k}(i)||$ to normalize the interpolator.
This procedure is based on the following result.

{\it Lemma:} Let ${\bf R}$ be a positive semi-definite Hermitian
symmetric matrix and ${\bf q}_{min}$ the eigenvector associated to
the smallest eigenvalue. If ${\bf q}_{min}$ is unique and of unit
norm, then with $\nu =1/tr[{\bf R}]$ the sequence of vectors ${\bf
v}(i)=\frac{\hat{\bf v}(i)}{||\hat{\bf v}(i)||}$ with $\hat{\bf
v}(i)= ({\bf I}-\nu(i){\bf R})\hat{\bf v}(i-1)$ converges to ${\bf
q}_{min}$, provided that $\hat{\bf v}(0)$ is not orthogonal to
${\bf q}_{min}$. A proof is shown in the Appendix.

To recursively estimate the matrix $\bar{\bf R}_{k}(i)$ and avoid
its inversion we use the matrix inversion lemma and Kalman RLS
recursions \cite{haykin}:
\begin{equation}
{\bf G}(i) =  \frac{\alpha^{-1}{\hat{\bar{\bf
R}}}_{k}^{-1}(i-1){\bar{\bf r}}_{k}(i)} {1+ \alpha^{-1} {\bar{\bf
r}}^{H}_{k}(i) {\hat{\bar{\bf R}}}_{k}^{-1}(i-1) {\bar{\bf
r}}_{k}(i)}
\end{equation}
\begin{equation}
{\hat{\bar{\bf R}}}_{k}^{-1}(i) = \alpha^{-1} {\hat{\bar{\bf
R}}}_{k}^{-1}(i-1) -\alpha^{-1}{\bf G}(i) {\bar{\bf r}}^{H}_{k}(i)
{\hat{\bar{\bf R}}}_{k}^{-1}(i-1)
\end{equation}
where $0<\alpha\leq 1$ is the forgetting factor. The algorithm can
be initialized with { ${\bar{\bf R}}_{k}^{-1}(0)=\delta {\bf I}$}
and ${\bf R}^{-1}_{{\bf u}_{k}}(0)=\delta {\bf I}$, where $\delta$
is a large positive number. For the computation of the
reduced-rank receiver parameter vector ${\bf w}_{k}$ we use the
matrix inversion lemma \cite{haykin} to estimate $({\bf
C}_{k}^{H}{\bf D}^{H}\bar{\bf R}_{k}^{-1}(i){\bf DC}_{k})^{-1}$ as
given by:
\begin{equation}
\boldsymbol{\Gamma}^{-1}_{k}(i) = \frac{1}{1-\alpha}\Bigg[
\boldsymbol{\Gamma}^{-1}_{k}(i-1)-
\frac{\boldsymbol{\Gamma}^{-1}_{k}(i-1)
\boldsymbol{\gamma}_{k}(i)\boldsymbol{\gamma}^{H}_{k}(i)\boldsymbol{\Gamma}^{-1}_{k}(i-1)}
{\frac{1-\alpha}{\alpha}+\boldsymbol{\gamma}^{H}_{k}(i)\boldsymbol{\Gamma}^{-1}_{k}(i)\boldsymbol{\gamma}_{k}(i)}\Bigg]
 \end{equation} where
$\boldsymbol{\Gamma}_{k}(i)$ is an estimate of $({\bf
C}_{k}^{H}{\bf D}^{H}{\hat{\bar{\bf R}}}^{-1}_{k}(i){\bf D}{\bf
C}_{k})$ and $\boldsymbol{\gamma}_{k}(i)={\bf C}_{k}^{H}{\bf
D}^{H}{\bf r}_{k}(i)$ and then we construct the reduced-rank
receiver as :
\begin{equation}
{\bf w}_{k}(i) = {\hat{\bar{\bf R}}}_{k}(i)^{-1} {\bf DC}_{k}
\boldsymbol{\Gamma}^{-1}_{k}(i){\hat{\bf g}}(i)
\end{equation}

The channel estimate $\hat{\bf g}(i)$ is obtained through the
power method and the RLS technique described in \cite{power2}.
Following this approach, the SVD on the $L_{p}\times L_{p}$ matrix
${\bf C}_{k}^{H}{\bf R}^{-1}(i){\bf C}_{k}$, as stated in (19) and
that requires $O(L_{p}^3)$, is avoided and replaced by a single
matrix-vector multiplication, resulting in the reduction of the
corresponding computational complexity on one order of magnitude
and no performance loss. In terms of computational complexity, the
CMV-RLS algorithm with the interpolated  receiver trades off one
blind algorithm with complexity {  $O(M^2)$} against two with
complexity { $O(M^2/L^2)$} and { $O(N_{I}^{2})$} operating in
parallel. Since { $N_{I}$} is small as compared to {$M$}, it turns
out that the new algorithms offer a significant computational
advantage over conventional RLS algorithms.

\subsection{Computational Complexity}

In this section we illustrate the computational complexity of the
proposed INT structure and algorithms. In Table 1 we consider
supervised algorithms, whereas the complexity of blind algorithms
is depicted in Table 2. Specifically, we compare the full-rank,
the proposed INT structure, the PD, the PC and the MWF structures
with SG and RLS algorithms.

In general, the INT structure introduces the term $M/L$, which can
reduce the complexity by choosing the decimation factor $L \geq
2$. This is relevant for algorithms which have quadratic
computational cost with $M$, i.e. the blind and trained RLS and
the blind SG, because the decimation factor $L$ in the denominator
favors the proposed scheme which requires complexity $O((M/L)^2)$.
This complexity advantage is not verified with linear complexity
recursions. For instance, with NLMS algorithms the proposed INT
has a complexity slightly superior to the full-rank. Among the
other methods, the PD is slightly more complex than the INT. A
drawback of PC methods is that they require an SVD with associated
cost $O(M^{3})$ in order to compute the desired subspace. Although
the subspace of interest can be retrieved via computationally
efficient tracking algorithms \cite{wang&poor,song&roy}, these
algorithms are still complex ($O(M)^2$) and lead to performance
degradation as compared to the actual SVD. The MWF technique has a
complexity $O(D \bar{M}^{2})$, where the variable dimension of the
vectors $\bar{M} = M - d$ varies according to the orthogonal
decomposition and the rank $d = 1, \ldots, D$.

\begin{figure}[!htb]
\begin{center}
\def\epsfsize#1#2{1\columnwidth}
\epsfbox{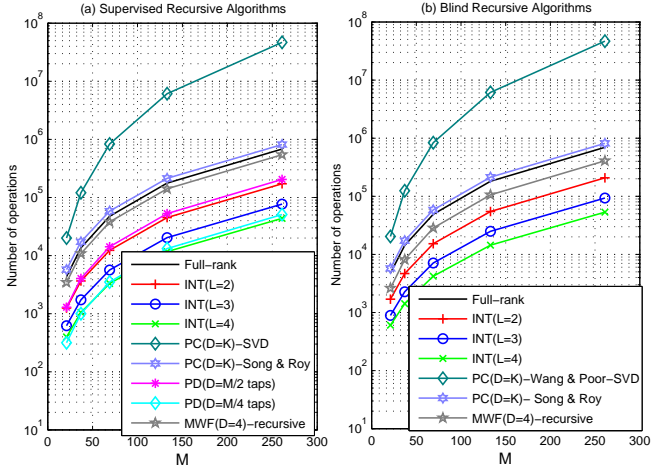} \caption{\small Complexity in terms of
arithmetic operations versus number of received samples (M) for
(a) supervised and (b) blind recursive adaptation algorithms.}
\end{center}
\end{figure}

In order to illustrate the complexity trend in a comprehensive
way, we depict in Fig. 2 curves which describe the computational
complexity in terms of the arithmetic operations (additions and
multiplications) as a function of the number of parameters $M$ for
recursive algorithms. For these curves, we consider $L_{p}=6$ and
assume that $D$ is equal to $M/2$ for the eigen-decomposition
approaches. We also include the computational cost of the
algorithm of Song and Roy \cite{song&roy}, which is capable of
significantly reducing the cost required by SVD.  In comparison
with the existing reduced-rank techniques, the proposed INT scheme
is significantly less complex than the PC and the MWF and slightly
less complex than the PD. This is because the analyzed algorithms
have quadratic cost (PC with SVD has cubic cost), whereas the INT
has complexity $O((M/L)^2)$, as shown in Tables I and II.


\begin{table}[h]
\centering%
\caption{\small Computational complexity of supervised adaptation
algorithms.} {
\begin{tabular}{ccc}
\hline \rule{0cm}{2.5ex}&  \multicolumn{2}{c}{Number of operations
per symbol } \\ \cline{2-3}
Algorithm & Additions & Multiplications \\
\hline
\emph{\small \bf LMS-Full-rank} & {\small $2M$} & {\small $2M+1$}  \\
\emph{\small \bf LMS-INT}   & {\small $2\frac{M}{L}+2N_{I}$} & {\small $3\frac{M}{L} +2N_{I}$}  \\
\emph{}   & {\small $+ N_{I}M +(\frac{M}{L})N_{I}+2$} & {\small $+(\frac{M}{L})N_{I}$}  \\
\emph{\small \bf LMS-PC} & {\small $O(M^{3}) + 2D$} & {\small $M^{3} + 2D +1$}\\
\emph{\small \bf LMS-PD} & {\small $(D-1)^{2} + 2D +1$} & {\small $D^{2} + 2D + 2$}  \\
\emph{\small \bf MWF-SG} & {\small $D(2(\bar{M}-1)^{2} +\bar{M} + 3)$} & {\small $D(2\bar{M}^{2} +5\bar{M} + 7)$}  \\
\hline
\emph{\small \bf RLS-Full-rank} & {\small $3(M-1)^{2} + M^{2} + 2M$} & {\small $6M^{2}+2M + 2$}   \\
\emph{}  & {\small $3(\frac{M}{L}-1)^{2}+3(N_{I}-1)^{2}$} & {\small $6(\frac{M}{L})^{2}+6N_{I}^{2}$} \\
\emph{\small \bf RLS-INT}  & {\small $+(\frac{M}{L}-1)N_{I}$} & {\small $\frac{M}{L}N_{I}$} \\
\emph{}  & {\small $+ N_{I}M+(\frac{M}{L})^{2}$} & {\small $+3\frac{M}{L}$} \\
\emph{}  & {\small $+N_{I}^{2}+2\frac{M}{L}+2N_{I}$} & {\small $+N_{I}+2$} \\
\emph{\small \bf RLS-PC} & {\small $M^{3} + 3(D-1)^{2} $} & {\small $O(M^{3}) + 6D^{2}$}   \\
\emph{} & {\small $ + D^{2} + 2D$} & {\small $+2D + 2$}   \\
\emph{\small \bf RLS-PD}  & {\small $4(D-1)^{2} + D^{2} + 2D$} & {\small $7D^{2}+2D + 2$} \\
\emph{\small \bf MWF-Recursive}  & {\small $D(4(\bar{M}-1)^{2} + 2\bar{M})$} & {\small $D(4\bar{M}^{2} +2\bar{M} + 3)$} \\
\hline
\end{tabular}
}
\end{table}

\begin{table}[h]
\centering%
\caption{\small Computational complexity of blind adaptation
algorithms.} {
\begin{tabular}{ccc}
\hline \rule{0cm}{2.5ex}&  \multicolumn{2}{c}{Number of operations
per symbol } \\ \cline{2-3}
Algorithm & Additions  & Multiplications \\
\hline
\emph{\small \bf CMV-SG-} & {\small $M^{2} + ML_{p} + 2M +1$} & {\small $M^{2}+ML_{p} + 3M$}  \\
\emph{\small \bf Full-rank}   & {\small $(\frac{M}{L})^{2} + (\frac{M}{L})L_{p} $} & {\small $(\frac{M}{L})^{2} + (\frac{M}{L})L_{p} $}  \\
\emph{\small \bf CMV-SG-INT}   & {\small $ +N_{I}M + (\frac{M}{L})N_{I}$} & {\small $+ \frac{M}{L}N_{I}+4\frac{M}{L}$}  \\
\emph{}   & {\small $+2\frac{M}{L}+N_{I}+2$} & {\small $+N_{I}$}  \\

\emph{\small \bf MWF-SG} & {\small $D(2(\bar{M}-1)^{2} +2)$} & {\small $D(2\bar{M}^{2} +3\bar{M} + 5)$} \\
\hline
\emph{}  & {\small $4(M-1)^{2}+M^{2}$} & {\small $7M^{2}+M$} \\
\emph{\small \bf CMV-RLS-}  & {\small $+3(L_{p}-1)^{2}-1$} & {\small $+L_{p}^{2}$} \\
\emph{\small \bf Full-rank}   & {\small $L_{p}^{2}+2L_{p}+ML_{p}$} & {\small $+ML_{p}+L_{p}+4$} \\
\emph{} & {\small $4(\frac{M}{L}-1)^{2}+(\frac{M}{L})^{2}+L_{p}^{2}$} & {\small $7(\frac{M}{L})^{2}+2\frac{M}{L}$} \\
\emph{\small \bf CMV-RLS-INT}  & {\small $+3(L_{p}-1)^{2}+2\frac{M}{L}L_{p}$} & {\small $+L_{p}^{2}+\frac{M}{L}L_{p}$} \\
\emph{} & {\small $+N_{I}M+3L_{p}-1$} & {\small $+L_{p}+2+N_{I}^2$} \\
\emph{\small }  & {\small $ + (\frac{M}{L} -1)N_{I}+(N_{I}-1)^{2} $} & {\small $+\frac{M}{L}N_{I}+N_{I}$} \\
\emph{\small \bf PC-Wang\&Poor} & {\small $M^{3} + 2(M-1)^{2}$} & {\small $O(M^{3}) + 2M^{2} + M $}   \\
\emph{\small \bf MWF-Recursive}  & {\small $D(3(\bar{M}-1)^{2} + 2\bar{M})$} & {\small $D(3\bar{M}^{2} +2\bar{M} + 3)$} \\
\hline
\end{tabular}
}
\end{table}

\section{Global Convergence Properties of the Method and Convergence Analysis of Algorithms}

In this section we discuss the global convergence of the method
and its properties, the trajectory of the mean tap vectors, of the
excess mean square error and the convergence speed. Specifically,
we study the convergence properties of the proposed joint adaptive
interpolator and receiver scheme and conclude that it leads to an
optimization problem with multiple global minima and no local
minima. In this regard and based on the analyzed convergence
properties of the method, it suffices to examine the MSE
trajectory of only one of the jointly optimized parameter vectors
(${\bf w}_{k}$ or ${\bf v}_{k}$) in order to predict the excess
MSE of both blind and supervised adaptive algorithms. We also
provide a discussion of the speed of convergence of the INT as
compared to the full-rank.

\subsection{Global Convergence of the Method and its Properties}

\subsubsection{Interpolated MMSE Design}

Let us first consider the trained receiver case and recall the
associated MSE expressions in (13) and (14), namely $J_{MSE}({\bf
v}_{k},\boldsymbol{\alpha}({\bf v}_{k})) = J({\bf v}_{k})=
\sigma^{2}_{b_{k}} - \bar{\bf p}^{H}_{k}(i)\bar{\bf R}^{-1}_{k}(i)
\bar{\bf p}_{k}(i)$ and $J_{MSE}(\boldsymbol{\beta}({\bf
w}_{k}),{\bf w}_{k}) = \sigma^{2}_{b_{k}} - \bar{\bf
p}^{H}_{u_{k}}(i)\bar{\bf R}^{-1}_{u_{k}}(i) {\bf p}_{u_{k}}(i)$
where $\sigma^{2}_{b_{k}}=E[|b_{k}(i)|^{2}]$. Note that points of
global minimum of $J_{MSE}({\bf w}_{k}(i),{\bf v}_{k}(i)) =
E\Big[|b_{k}(i)- {\bf v}^{H}_{k}(i) \boldsymbol{\Re}(i){\bf
w}_{k}^{*}(i) |^{2}\Big]$ can be obtained by ${\bf v}_{opt}= \arg
\min_{{\bf v}_{k}}~ J({\bf v}_{k})$ and ${\bf
w}_{opt}=\boldsymbol{\alpha}({\bf v}_{opt})$ or ${\bf w}_{opt}=
\arg \min_{{\bf w}_{k}} J_{MSE}(\boldsymbol{\beta}({\bf
w}_{k}),{\bf w}_{k})$ and ${\bf v}_{opt}=\boldsymbol{\beta}({\bf
w}_{opt})$. At a minimum point $J_{MSE}({\bf
v}_{k},\boldsymbol{\alpha}({\bf v}_{k}))$ equals
$J_{MSE}(\boldsymbol{\beta}({\bf w}_{k}),{\bf w}_{k})$ and the
MMSE for the proposed structure is achieved. We further note that
since $J({\bf v}_{k})=J(t{\bf v}_{k})$, for every $t\neq0$, then
if ${\bf v}^{\star}_{k}$ is a point of global minimum of $J({\bf
v}_{k})$ then $t{\bf v}^{\star}_{k}$ is also a point of global
minimum. Therefore, points of global minimum (optimum interpolator
filters) can be obtained by ${\bf v}^{\star}_{k}=\arg \min_{||{\bf
v}_{k}||=1} J({\bf v}_{k})$. Since the existence of at least one
point of global minimum of $J({\bf v}_{k})$ for $||{\bf
v}_{k}||=1$ is guaranteed by the theorem of Weierstrass
\cite{bert} (see Chapter 2, Section 2.1, Appendix B), then the
existence of (infinite) points of global minimum is also
guaranteed for the cost function in (10).

In the context of global convergence, a sufficient but not
necessary condition is the convexity, which is verified if its
Hessian matrix is positive semi-definite, that is ${\bf a}^{H}{\bf
H}{\bf a} \geq 0$, for any vector ${\bf a}$. Firstly, let us
consider the minimization of $J_{MSE}({\bf w}_{k}(i),{\bf
v}_{k}(i)) = E\Big[|b_{k}(i)- {\bf v}^{H}_{k}(i)
\boldsymbol{\Re}(i){\bf w}_{k}^{*}(i) |^{2}\Big]$ with fixed
interpolators. Such optimization leads to the following Hessian
${\bf H} = \frac{\partial}{\partial{\bf w}^{H}_{k}}\frac{
(J_{MSE}(.))}{\partial{\bf w}_{k}}=E[{\bf r}_{k}(i){\bf
r}_{k}^{H}(i)]={\bf R}_{k}(i)$, which is positive semi-definite
and ensures the convexity of the cost function for the case of
fixed interpolators. Let us now consider the joint optimization of
the interpolator ${\bf v}_{k}$ and receiver ${\bf w}_{k}$ through
an equivalent cost function to (10):
\begin{equation}
\tilde{J}_{MSE}({\bf z}) = E[|b- {\bf z}^{H}_{k}{\bf B}{\bf
z}_{k}|^{2}]
\end{equation}
where  ${\bf B} = \left[\begin{array}{c c} {\bf 0} & {\bf 0}  \\
\boldsymbol{\Re }  & {\bf 0}  \\ \end{array}\right]$ is an
$(N_{I}+N/L) \times (N_{I}+N/L)$ matrix, the Hessian (${\bf H}$)
with respect to ${\bf z}_{k} = [{\bf w}^{T}_{k}~{\bf
v}^{T}_{k}]^{T}$ is ${\bf H} = \frac{\partial }{\partial{\bf
z}^{H}_{k}}\frac{\partial (\tilde{J}_{MSE}(.)) }{\partial{\bf
z}_{k}}= E[({\bf z}^{H}_{k}{\bf B}{\bf z}_{k} -b_{k}){\bf B}^{H}]
+ E[({\bf z}^{H}_{k}{\bf B}^{H}{\bf z}_{k} -b^{*}_{k}){\bf
B}]+E[{\bf B}{\bf z}_{k}{\bf z}^{H}_{k}{\bf B}^{H}] + E[{\bf
B}^{H}{\bf z}_{k}{\bf z}^{H}_{k}{\bf B}]$. By examining ${\bf H}$
we note that the third and fourth terms yield positive
semi-definite matrices (${\bf a}^{H} E[{\bf B}{\bf z}_{k}{\bf
z}^{H}_{k}{\bf B}^{H}]{\bf a} \geq 0$ and ${\bf a}^{H} E[{\bf
B}^{H}{\bf z}_{k}{\bf z}^{H}_{k}{\bf B}]{\bf a} \geq 0$, ${\bf
z}_{k}\neq{\bf 0}$) whereas the first and second terms are
indefinite matrices. Thus, the cost function cannot be classified
as convex. However, for a gradient search algorithm, a desirable
property of the cost function is that it shows no points of local
minimum, i.e. every point of minimum is a point of global minimum
(convexity is a sufficient, but not necessary, condition for this
property to hold) and it is conjectured that the problem in (40)
has this property.

\begin{figure}[!htb]
\begin{center}
\def\epsfsize#1#2{1.0\columnwidth}
\epsfbox{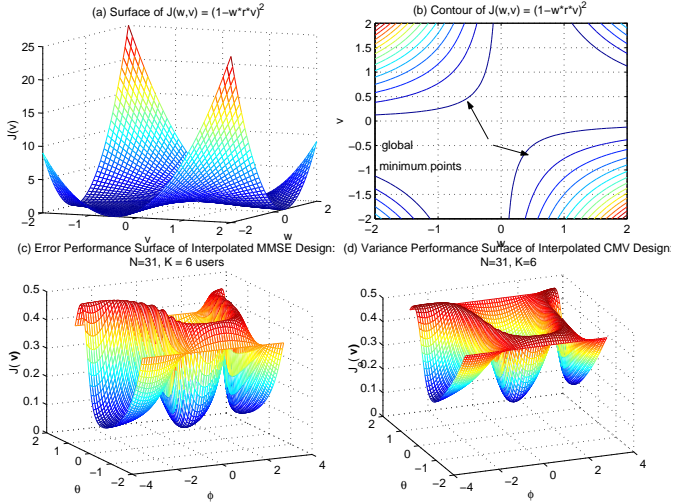} \caption{\small (a) Error Performance Surface
of the function $f(v,w)=(1-w*r*v)^{2}$, (b) Contour plots showing
that the function does not exhibit local minima and has multiple
global minima, (c) Error performance surface of interpolated MMSE
receivers at $E_{b}/N_{0}=15 dB$ for $L=3$ and(d) Variance
performance surface of $J_{MV}({\bf v})$ for CMV receivers at
$E_{b}/N_{0}=15 dB$ for $L=2$ and channel with paths given by $0$,
$-6$ and $-10$ dB, spaced by $T_{c}$.}
\end{center}
\end{figure}

To support this claim, we carried out the following studies:

\begin{itemize}

\item{Let us consider the scalar case of the function in (40),
which is defined as $f(w,v) = (b-w ~r~ v)^{2}=b^{2} - 2b~w ~r~v +
(w~\Re~ v)^{2}$, where $r$ is a constant. By choosing $v$ (the
"scalar" interpolator) fixed, it is evident that the resulting
function $f(w,v) = (b-w~c)^{2}$, where $c$ is a constant is a
convex one, whereas for a time-varying interpolator the curves
shown in Fig. 3 (a) and (b), indicate that the function is no
longer convex but it also does not exhibit local minima.}

\item{By taking into account that for small interpolator filter
length $N_{I}$ ($N_{I}\leq 3$), ${\bf v}_{k}$ can be expressed in
spherical coordinates and a surface  can be constructed.
Specifically, we expressed the parameter vector ${\bf v}_{k}$ as
follows: ${\bf v}_{k}= r [cos(\theta)cos(\phi)
~~~cos(\theta)sin(\phi) ~~~ sin(\theta)]^{T}$, where $r$ is the
radius, $\theta$ and $\phi$ were varied from $-\pi/2$ to $\pi/2$
and $-\pi$ to $\pi$, respectively, and (13) was plotted for
various scenarios and conditions (SNR, different channels, etc).
The plot of the error-performance surface of $J({\bf v}_{k})$,
depicted in Fig. 3 (c), reveals that $J({\bf v}_{k})$ has a global
minimum value (as it should) but do not exhibit local minima,
which implies that (40) has no local minima either. It should be
noted that if the cost function in (40) had a point of local
minimum then $J({\bf v}_{k})$ in (13) should also exhibit a point
of local minimum even though the reciprocal is not necessarily
true: a point of local minimum of $J({\bf v}_{k})$ may correspond
to a saddle point of $J_{MSE}({\bf v}_{k},{\bf w}_{k})$, if it
exists. Note also that the latitude X longitude plot in Fig. 3 (c)
depicts its two symmetric global minima in the unit sphere.}

\item{An important feature that advocates the
non-existence of local minima is that the algorithm always
converge to the same minimum value, for a given experiment,
independently of any interpolator initialization (except for ${\bf
v}(0)=[0 ~\dots~ 0]^{T}$ that eliminates the signal)  for a wide
range of SNR values and channels.}

\end{itemize}

\subsubsection{Interpolated CMV Design}

For the blind case, let us first consider the minimization of
$J_{MV}({\bf w}_{k}(i),{\bf v}_{k}(i)) = E\Big[|{\bf v}^{H}_{k}(i)
\boldsymbol{\Re}(i){\bf w}_{k}^{*}(i) |^{2}\Big]$ with fixed
interpolators subject to { ${\bf C}_{k}^{H}{\bf D}^{H} {\bf
w}_{k}(i) = {\bf g}(i)$} and {$||{\bf v}_{k}(i)||=1$}. It should
be noted that global convergence of the CMV method has been
established in \cite{xu&tsatsanis} and here we treat a similar
problem when fixed interpolators are used. Such optimization leads
to the following Hessian ${\bf H} = \frac{\partial}{\partial{\bf
w}^{H}_{k}}\frac{ (J_{MV}(.))}{\partial{\bf w}_{k}}=E[{\bf
r}_{k}(i){\bf r}_{k}^{H}(i)]={\bf R}_{k}(i)$, which is positive
semi-definite and ensures the convexity of the cost function for
the case of fixed interpolators.

Consider the joint optimization of the interpolator ${\bf v}_{k}$
and receiver ${\bf w}_{k}$ via an equivalent cost function to
(10):
\begin{equation}
\tilde{J}_{MV}({\bf z}) = E[|{\bf z}^{H}_{k}{\bf B}{\bf
z}_{k}|^{2}]
\end{equation}
subject to ${\bf C}_{k}^{H}{\bf D}^{H} {\bf w}_{k}(i) = {\bf
g}(i)$, where  ${\bf B} = \left[\begin{array}{c c} {\bf 0} & {\bf 0}  \\
\boldsymbol{\Re }  & {\bf 0}  \\ \end{array}\right]$ is an
$(N_{I}+N/L) \times (N_{I}+N/L)$ matrix, the Hessian (${\bf H}$)
with respect to ${\bf z}_{k} = [{\bf w}^{T}_{k}~{\bf
v}^{T}_{k}]^{T}$ is ${\bf H} = \frac{\partial }{\partial{\bf
z}^{H}_{k}}\frac{\partial (\tilde{J}_{MSE}(.)) }{\partial{\bf
z}_{k}}= E[{\bf z}^{H}_{k}{\bf B}{\bf z}_{k}{\bf B}^{H}] + E[{\bf
z}^{H}_{k}{\bf B}^{H}{\bf z}_{k}{\bf B}]+E[{\bf B}{\bf z}_{k}{\bf
z}^{H}_{k}{\bf B}^{H}] + E[{\bf B}^{H}{\bf z}_{k}{\bf
z}^{H}_{k}{\bf B}]$. By examining ${\bf H}$ we note that, as
occurs for the MMSE case, the third and fourth terms yield
positive semi-definite matrices (${\bf a}^{H} E[{\bf B}{\bf
z}_{k}{\bf z}^{H}_{k}{\bf B}^{H}]{\bf a} \geq  0$ and ${\bf a}^{H}
E[{\bf B}^{H}{\bf z}_{k}{\bf z}^{H}_{k}{\bf B}]{\bf a} \geq 0$,
${\bf z}_{k}\neq{\bf 0}$) whereas the first and second terms are
indefinite matrices. Hence, the cost function cannot be classified
as convex, although we conjecture that it does not exhibit local
minima. Thus, we proceed similarly to the MMSE case to study the
surfaces provided by the problem in (41). Then, we carried out the
following studies:

\begin{itemize}

\item{We have also plotted the variance performance surface of
$J_{o}({\bf v}_{k})$ in (17), depicted in Fig. 3 (d). This surface
reveals that $J_{o}({\bf v}_{k})$ has a global minimum (as it
should) but does not exhibit local minima, which implies that (41)
subject to ${\bf C}_{k}^{H}{\bf D}^{H} {\bf w}_{k}(i) = {\bf
g}(i)$ has no local minima either. }

\item{Another important feature that suggests the
non-existence of local minima for the blind algorithms is that
they always converge to the same minimum value, for a given
experiment, independently of any interpolator initialization
(except for ${\bf v}(0)=[0 ~\dots~ 0]^{T}$ that eliminates the
signal) for a wide range of parameters.}

\end{itemize}

\subsection{Trajectory of the Mean Tap Vectors}

This part is devoted to the analysis of the trajectory of the mean
tap vectors of the proposed structure when operating in blind and
supervised modes. In our analysis, we employ the so called
Independence Theory \cite{1,haykin} (see Chapter $9$, pp. 390-404)
that consists of four points, namely:

1. The received vectors ${\bf r}(1), \ldots, {\bf r}(i)$ and their
interpolated counterparts $\bar{\bf r}_{k}(1), \ldots, \bar{\bf
r}_{k}(i)$ constitute a sequence of statistically independent
vectors.

2. At time $i$, ${\bf r}(i)$ and $\bar{\bf r}_{k}(i)$ are
statistically independent of $b_{k}(1), \ldots, b_{k}(i-1)$.

3. At time $i$, $b_{k}(i)$ depends on ${\bf r}(i)$ and ${\bf
r}_{k}(i)$, but is independent of previous $b_{k}(n)$, for $n=1,
\ldots, i-1$.

4. The vectors ${\bf r}(i)$ and $\bar{\bf r}_{k}(i)$ and the
sample $b_{k}$ are mutually Gaussian-distributed random variables.

In the present context, it is worth noting that the independence
assumption holds for synchronous DS-CDMA systems \cite{1}, which
is the present case, but not for asynchronous models, even though
it provides substantial insight.

\subsubsection{Trained algorithm}

To proceed, let us drop the user $k$ index for ease of
presentation and define the tap error vectors ${\bf e}_{w}(i)$ and
${\bf e}_{v}(i)$ at time index $i$
\begin{equation}
{\bf e}_{w}(i) = {\bf w}(i) - {\bf w}_{opt},~{\bf e}_{v}(i) = {\bf
v}(i) - {\bf v}_{opt}
\end{equation}
where ${\bf w}_{opt}$ and ${\bf v}_{opt}$ are the optimum tap
vectors that achieve the MMSE for the proposed structure.
Substituting the expressions in (42) into (21) and (22) we get
\begin{equation}
{\bf e}_{w}(i+1) = [{\bf I} - \mu \bar{\bf r}(i)\bar{\bf
r}^{H}(i)] {\bf e}_{w}(i) + \mu \bar{\bf r}(i) e^{*}(i)
\end{equation}
\begin{equation}
{\bf e}_{v}(i+1) = [{\bf I} - \eta {\bf u}(i){\bf u}^{H}(i)] {\bf
e}_{v}(i) + \eta {\bf u}(i) e^{*}(i)
\end{equation}
By taking expectations on both sides we have
\begin{equation}
E[{\bf e}_{w}(i+1)] = [{\bf I} - \mu \bar{\bf R}(i)] E[{\bf
e}_{w}(i)] + \mu E[\bar{\bf r}(i) e^{*}(i)]
\end{equation}
\begin{equation}
E[{\bf e}_{v}(i+1)] = [{\bf I} - \eta {\bf R}_{\bf u}(i)] E[{\bf
e}_{v}(i)] + \eta E[{\bf u}(i) e^{*}(i)]
\end{equation}
At this point, it should be noted that the two error vectors have
to be considered together because of the joint optimization of the
interpolator filter and the reduced-rank filter. Rewriting the
terms $E[\bar{\bf r}(i) e^{*}(i)]$ and $E[{\bf u}(i) e^{*}(i)]$,
using (42) and the independence theory \cite{haykin} (see Chap.
$9$, pp. 390-404) we obtain
\begin{equation}
\begin{split}
E[\bar{\bf r}(i) e^{*}(i)] & = \bar{\bf p}(i) - E[\bar{\bf
r}(i){\bf v}^{T}(i)\boldsymbol{\Re}^{H}(i)]E[{\bf e}_{w}(i)]- \\ &
E[\bar{\bf r}(i){\bf w}_{opt}^{T}\boldsymbol{\Re}^{*}]E[{\bf
e}_{v}(i)] - E[\bar{\bf r}(i) {\bf
w}_{opt}^{T}\boldsymbol{\Re}^{*}{\bf v}_{opt}]
\end{split}
\end{equation}
\begin{equation}
\begin{split}
E[{\bf u}(i) e^{*}(i)] & = \bar{\bf p}_{\bf u}(i) - E[{\bf
u}(i){\bf w}^{T}(i)\boldsymbol{\Re}^{*}(i)]E[{\bf e}_{v}(i)]\\ &
-E[{\bf u}(i){\bf v}_{opt}^{T}\boldsymbol{\Re}^{H}]E[{\bf
e}_{w}(i)] -E[{\bf u}(i) {\bf w}_{opt}^{T}\boldsymbol{\Re}^{*}{\bf
v}_{opt}]
\end{split}
\end{equation}
By combining (45), (46), (47) and (48) the trajectory of the error
vectors is given by:
\begin{equation}
\left[\begin{array}{c}
  E[{\bf e}_{w}(i+1)] \\
  E[{\bf e}_{v}(i+1)]
\end{array}\right] = {\bf A}
\left[\begin{array}{c}
  E[{\bf e}_{w}(i)] \\
  E[{\bf e}_{v}(i)]
\end{array}\right] + {\bf B}
\end{equation}
where \\ {\footnotesize ${\bf A}
=\left[\hspace*{-0.8em}\begin{array}{c c}
  ({\bf I} - \mu \bar{\bf R}) - \mu E[\bar{\bf r}(i){\bf v}^{T}(i)\boldsymbol{\Re}^{H}(i)] & \hspace*{-1em} - \mu E[\bar{\bf r}(i){\bf w}_{opt}^{T}\boldsymbol{\Re}^{*}(i)] \\
- \eta E[{\bf u}(i){\bf v}_{opt}^{T}\boldsymbol{\Re}^{H}(i)] &
\hspace*{-1em} ({\bf I} - \eta \bar{\bf R}_{\bf u}) - \eta E[{\bf
u}(i){\bf w}^{T}(i)\boldsymbol{\Re}^{*}(i)]
\end{array}\hspace*{-0.8em}\right]$} and \\ ${\bf B} = \left[\begin{array}{c}
  \mu \bar{\bf p}(i) - \mu E[\bar{\bf r}(i) {\bf w}_{opt}^{T}\boldsymbol{\Re}^{*}{\bf
v}_{opt}] \\
  \eta \bar{\bf p}_{\bf u}(i) -\eta E[{\bf u}(i) {\bf w}_{opt}^{T}\boldsymbol{\Re}^{*}{\bf v}_{opt}]
\end{array}\right]$. Equation (49) implies that the stability
of the algorithms in the proposed structure depends on the matrix
${\bf A}$. For stability, the convergence factors should be chosen
so that the eigenvalues of ${\bf A}^{H}{\bf A}$ are less than one.

\subsubsection{Blind algorithm}

The mean vector analysis of the blind algorithm is slightly
different from \cite{xu&tsatsanis} because our approach uses a
decoupled SG channel estimation technique \cite{power2}, that
yields better channel estimates. Hence, we consider the joint
estimation of ${\bf w}_{k}$ and ${\bf v}_{k}$, while ${\bf g}$ is
a decoupled estimation process. To proceed, let us drop the user
$k$ index for ease of presentation and substitute the expressions
of (42) into (32) and (34) that gives:
\begin{equation}
\begin{split}
{\bf e}_{w}(i+1) & = [{\bf I} - \mu \bar{\bf r}(i)\bar{\bf
r}^{H}(i)] {\bf e}_{w}(i) + {\bf D C}({\bf C}^{H}{\bf D}^{H}{\bf D
C })^{-1}{\bf g}(i) \\ & - \mu \boldsymbol{\Pi}\bar{\bf r}(i){\bf
v}_{opt}^{T}\boldsymbol{\Re}^{*}(i) {\bf w}_{opt} - \mu
\boldsymbol{\Pi}\bar{\bf r}(i){\bf
w}_{opt}^{T}\boldsymbol{\Re}^{H}(i){\bf e}_{v}(i)
\end{split}
\end{equation}
\begin{equation}
\begin{split}
{\bf e}_{v}(i+1) & = [{\bf I} - \eta {\bf u}(i){\bf u}^{H}(i)]
{\bf e}_{v}(i) - \eta {\bf u}(i) {\bf
v}_{opt}^{T}\boldsymbol{\Re}^{*}(i){\bf e}_{w}(i) \\ & - \eta {\bf
u}(i) {\bf w}_{opt}^{T} \boldsymbol{\Re}^{*}(i) {\bf v}_{opt}
\end{split}
\end{equation}
where $\boldsymbol{\Pi} = {\bf I} -{\bf DC}({\bf C}^{H} {\bf
D}^{H}{\bf D}{\bf C})^{-1}{\bf C}^{H}{\bf D}^{H}$ and we used the
fact that the scalars have alternative expressions as $({\bf
e}_{w}^{T}(i)\boldsymbol{\Re}^{H}(i){\bf v}_{opt})^{T} = ({\bf
e}_{w}^{T}(i)\boldsymbol{\Re}^{H}(i){\bf v}_{opt}) = {\bf
v}_{opt}^{T}\boldsymbol{\Re}^{*}(i){\bf e}_{w}(i)$ and $({\bf
e}_{v}^{T}(i)\boldsymbol{\Re}^{*}(i){\bf w}_{opt})^{T} = ({\bf
e}_{v}^{T}(i)\boldsymbol{\Re}^{*}(i){\bf w}_{opt}) = {\bf
w}_{opt}^{T}\boldsymbol{\Re}^{H}(i){\bf e}_{v}(i)$. By taking
expectations on both sides and eliminating the term $\mu
\boldsymbol{\Pi}\bar{\bf r}(i){\bf v}_{opt}\boldsymbol{\Re}^{*}(i)
{\bf w}_{opt}$ we get
\begin{equation}
\begin{split}
E[{\bf e}_{w}(i+1)] & = [{\bf I} - \mu \bar{\bf R}(i)] E[{\bf
e}_{w}(i)] + {\bf D C}({\bf C}^{H}{\bf D}^{H}{\bf D C
})^{-1}E[{\bf g}(i)] \\& - \mu \boldsymbol{\Pi}E[\bar{\bf
r}(i){\bf w}_{opt}^{T}\boldsymbol{\Re}^{H}(i)]E[{\bf e}_{v}(i)]
\end{split}
\end{equation}
\begin{equation}
\begin{split}
E[{\bf e}_{v}(i+1)] & = [{\bf I} - \eta {\bf R}_{\bf u}(i)] E[{\bf
e}_{v}(i)] - \eta E[{\bf u}(i) {\bf
v}_{opt}^{T}\boldsymbol{\Re}^{*}(i)]E[{\bf e}_{w}(i)] \\& - \eta
E[{\bf u}(i) {\bf w}_{opt}^{T} \boldsymbol{\Re}^{*}(i)] {\bf
v}_{opt}
\end{split}
\end{equation}
By combining (52) and (53) the trajectory of the error vectors for
the minimum variance case is given by:
\begin{equation}
\left[\begin{array}{c}
  E[{\bf e}_{w}(i+1)] \\
  E[{\bf e}_{v}(i+1)]
\end{array}\right] = {\bf A}_{MV}
\left[\begin{array}{c}
  E[{\bf e}_{w}(i)] \\
  E[{\bf e}_{v}(i)]
\end{array}\right] + {\bf B}_{MV}
\end{equation}
where ${\bf A}_{MV} =\left[\begin{array}{c c}
  [{\bf I} - \mu \bar{\bf R}(i)] & - \mu \boldsymbol{\Pi}E[\bar{\bf r}(i){\bf
w}_{opt}^{T}\boldsymbol{\Re}^{H}(i)] \\
- \eta E[{\bf u}(i) {\bf v}_{opt}^{T}\boldsymbol{\Re}^{*}(i)] &
[{\bf I} - \eta {\bf R}_{\bf u}(i)] ]
\end{array}\right]$ and  ${\bf B}_{MV} = \left[\begin{array}{c}
  {\bf D C}({\bf C}^{H}{\bf D}^{H}{\bf D C })^{-1}E[{\bf g}(i)] \\
  -\eta E[{\bf u}(i) {\bf w}_{opt}^{T} \boldsymbol{\Re}^{*}(i)]
  {\bf v}_{opt}
\end{array}\right]$. Equation (54) suggests that the stability
of the algorithms in the proposed structure depends on the matrix
${\bf A}_{MV}$. For stability, the convergence factors should be
chosen so that the eigenvalues of ${\bf A}^{H}_{MV}{\bf A}_{MV}$
are less than one.

\subsection{Trajectory of Excess MSE}

Here we describe the trajectory of the excess MSE at steady-state
of the trained and the blind SG algorithms.

\subsubsection{Trained Algorithm}

The analysis for the LMS algorithm using the proposed interpolated
structure and the computation of its steady-state excess MSE
resembles the one in \cite{haykin} (see Chapter $9$, pp. 390-404).
Here, an interpolated structure with joint optimization of
interpolator ${\bf v}_{k}$ and reduced-rank receiver ${\bf w}_{k}$
is taken into account. Despite the joint optimization, for the
computation of the excess MSE one has to consider only the
reduced-rank parameter vector ${\bf w}_{k}$ because the MSE
attained upon convergence by (13) and (14) should be the same.
Here, we will drop the user $k$ index for ease of presentation.
Consider the MSE at time $i+1$ as:
\begin{equation}
\epsilon(i+1) = E[|b(i+1) - {\bf w}^{H}(i+1)\bar{\bf r}(i+1)|^{2}]
\end{equation}
By using ${\bf w}(i+1)={\bf w}_{opt} + {\bf e}_{w}(i+1)$, ${\bf
w}_{opt}$, ${\bf v}_{opt}$ and the fact that the expressions in
(13) and (14) are equal for the optimal parameter vectors, the MSE
becomes
\begin{equation}
\begin{split}
\epsilon(i+1) & = \sigma_{b}^{2} - \bar{\bf p}^{H}(i+1)\bar{\bf
R}^{-1}(i+1)\bar{\bf p}(i+1) \\ &\quad  -\bar{\bf p}^{H}(i+1){\bf
e}_{w}(i+1) - {\bf e}_{w}^{H}(i+1)\bar{\bf p}(i+1) \\ &\quad -
{\bf w}_{opt}^{H}\bar{\bf p}(i+1)  + {\bf w}_{opt}^{H}\bar{\bf
R}(i+1){\bf w}_{opt} \\ &\quad + {\bf w}_{opt}^{H} \bar{\bf
R}(i+1) {\bf e}_{w}(i+1) + {\bf e}_{w}^{H}(i+1) \bar{\bf
R}(i+1){\bf w}_{opt}
\\ &\quad + E[{\bf e}_{w}(i+1)\bar{\bf r}(i+1)\bar{\bf
r}^{H}(i+1){\bf e}_{w}^{H}(i+1)] \\ &  = \sigma_{b}^{2} - \bar{\bf
p}^{H}(i+1)\bar{\bf R}^{-1}(i+1)\bar{\bf p}(i+1) + \\ & \quad
E[{\bf e}_{w}(i+1)\bar{\bf r}(i+1)\bar{\bf r}^{H}(i+1){\bf
e}_{w}^{H}(i+1)] \\ & = J_{MMSE}({\bf w}_{opt},{\bf v}_{opt}) +
\xi_{exc}(i+1)
\end{split}
\end{equation}
where $\bar{\bf p}(i+1)=E[b^{*}(i+1)\bar{\bf r}(i+1)]$,
$\epsilon_{min}=J_{MMSE}({\bf w}_{opt},{\bf
v}_{opt})=\sigma_{b}^{2} - \bar{\bf p}^{H}(i+1)\bar{\bf
R}^{-1}(i+1)\bar{\bf p}(i+1)$ is the MMSE achieved by the proposed
structure when we have ${\bf w}_{opt}$ and ${\bf v}_{opt}$ and
$\xi_{exc}(i+1) = E[{\bf e}_{w}^{H}(i+1) \bar{\bf r}(i+1)\bar{\bf
r}^{H}(i+1) {\bf e}_{w}(i+1)]$ is the excess MSE at time $i+1$. To
compute the excess MSE one must evaluate the term
$\xi_{exc}(i+1)$. By invoking the independence assumption and the
properties of trace \cite{haykin} (see Chap. $9$, pp. 390-404) we
may reduce it as follows:
\begin{equation}
E[{\bf e}_{w}^{H}(i+1) \bar{\bf r}(i+1)\bar{\bf r}^{H}(i+1) {\bf
e}_{w}(i+1)] = tr\Big[ \bar{\bf R}(i+1) {\bf K}(i+1)\Big]
\end{equation}
In the following steps, we assume that $i$ is sufficiently large
such that the matrix $\bar{\bf R}(i)=\bar{\bf R}(\infty)=\bar{\bf
R}$. To proceed let us define some new quantities that will
perform a rotation of coordinates to facilitate our analysis as
advocated in \cite{haykin}. Define ${\bf Q}^{H} \bar{\bf R} {\bf
Q}=\boldsymbol{\Lambda}$, where $\boldsymbol{\Lambda}$ is a
diagonal matrix consisting of the eigenvalues of $\bar{\bf R}$ and
${\bf Q}$ is the unitary matrix with the eigenvectors associated
with these eigenvalues. Letting ${\bf Q}^{H}{\bf K}{\bf Q}={\bf
X}$ we get
\begin{equation}
\begin{split}
\xi_{exc}(i+1) & = tr\Big[ \bar{\bf R} {\bf K}(i+1)\Big] =
tr\Big[{\bf Q}\boldsymbol{\Lambda} {\bf Q}^{H} {\bf Q} \bar{\bf
X}(i+1) {\bf Q}^{H}\Big] \\& = tr\Big[{\bf Q}\boldsymbol{\Lambda}
\bar{\bf X}(i+1) {\bf Q}^{H}\Big] = tr\Big[\boldsymbol{\Lambda}
\bar{\bf X}(i+1) \Big]
\end{split}
\end{equation}
where we used the property of trace and ${\bf Q}^{H}{\bf Q}={\bf
I}$. Because $\boldsymbol{\Lambda}$ is a diagonal matrix of
dimension $M/L$ we have
\begin{equation}
\xi_{exc}(i+1) = \sum_{n=1}^{M/L}\lambda_{n}x_{n}(i+1)
\end{equation}
where $x_{n},~n=1,2,~\ldots~M/L$ are the elements of the diagonal
of ${\bf X}(i)$. Here, we may use (45), invoke the independence
theory \cite{haykin} (see Chapter $9$, pp. 390-404) in order to
describe the correlation matrix of the weight error vector:
\begin{equation}
\begin{split}
{\bf K}(i+1) & =E[{\bf e}_{w}(i+1){\bf e}_{w}^{H}(i+1)] \\ & =
({\bf I} -\mu \bar{\bf R}(i)){\bf K}(i)({\bf I} -\mu \bar{\bf
R}(i)) + \mu^{2} \epsilon_{min}
\end{split}
\end{equation}
Next, using the transformations ${\bf Q}^{H} \bar{\bf R} {\bf
Q}=\boldsymbol{\Lambda}$, ${\bf Q}^{H}{\bf K}{\bf Q}={\bf X}$ and
similarly to \cite{haykin} (see Chapter $9$, pp. 390-404), a
recursive equation in terms of ${\bf X}(i)$ and
$\boldsymbol{\Lambda}$ can be written:
\begin{equation}
{\bf X}(i+1) = ({\bf I}-\mu \boldsymbol{\Lambda}){\bf X}(i)({\bf
I} - \mu \boldsymbol{\Lambda}) + \mu^{2} \epsilon_{min}
\boldsymbol{\Lambda}
\end{equation}
Because of the structure of the above equation, one can decouple
the elements $x_{n}(i)$ from the off-diagonal ones, and thus
$\xi_{exc}(i+1)$ depends on $x_{n}(i)$ according to the following
recursion:
\begin{equation}
{x}_{n}(i+1) = (1-\mu {\lambda}_{n})^{2}{x}_{n}(i) + \mu^{2}
\epsilon_{min}{\lambda}_{n}
\end{equation}
At this point, it can be noted that such recursive relation
converges provided that all the roots lie inside the unit circle,
i. e.,  $(1-\mu \lambda_{n})^{2} < 1$ for all $n$, and thus we
have for stability
\begin{equation}
0<\mu<\frac{2}{\lambda_{max}}
\end{equation}
where $\lambda_{max}$ is the largest eigenvalue of the matrix
$\bar{\bf R}$. In practice, $tr[\bar{\bf R}]$ is used as a
conservative estimate of $\lambda_{max}$. By taking $\lim_{i
\rightarrow \infty}$ on both sides of (62), we get $x_{n}(\infty)
= \frac{\mu}{2 + \mu \lambda_{n}} \epsilon_{min}$. Then, taking
limits on both sides of (59) and using $x_{n}(\infty)$ we obtain
the expression for the excess MSE at steady-state
\begin{equation}
\xi_{exc}(\infty) = \sum_{n=1}^{M/L}\lambda_{n} x_{n}(\infty) =
\sum_{n=1}^{M/L} \frac{\mu \lambda_{n}}{2 + \mu \lambda_{n}}
\epsilon_{min} = \frac{\frac{\mu}{2} ~tr[\bar{\bf R}]}{1 -
\frac{\mu}{2}~ tr[\bar{\bf R}]} \epsilon_{min}
\end{equation}
The expression in (64) can be used to predict semi-analytically
the excess MSE, where $\bar{\bf R}$ must be estimated with the aid
of computer simulations since it is a function of the interpolator
${\bf v}(i)$. Alternatively, one can conduct the analysis for the
interpolator ${\bf v}(i)$, which results in the expression
$\xi_{exc}(\infty) =  \frac{\frac{\eta}{2} ~tr[{\bf R}_{\bf u}]}{1
- \frac{\eta}{2}~ tr[{\bf R}_{\bf u}]} \epsilon_{min}$, where
$\eta$ is the step size of the interpolator, the matrix ${\bf
R}_{u}={\bf R}_{u}(\infty)$ and ${\bf R}_{u}(i) = E[{\bf u}(i){\bf
u}^{H}(i)]$, as defined in connection with (12). A more complete
analytical result, expressed as a function of both step sizes,
$\mu$ and $\eta$, and statistics of the non-interpolated
observation vector ${\bf r}(i)$ requires further investigation in
order to determine $tr[\bar{\bf R}(\infty)]$, that depends on
$\eta$ or $tr[{\bf R}_{\bf u}(\infty)]$, that depends on $\mu$.
Nevertheless, such investigation is beyond the scope of this paper
and it should be remarked that the results would not differ from
the semi-analytical results derived here (that implicitly take
into account the parameters of ${\bf v}(i)$).

\subsubsection{Blind Algorithm}

Our algorithm is a minimum variance technique and its steady-state
excess MSE resembles the approach in \cite{xu&tsatsanis}. In the
current context, however, an interpolated structure with joint
optimization of interpolator ${\bf v}_{k}$ and reduced-rank
receiver ${\bf w}_{k}$ is taken into account. In particular, it
suffices to consider for the computation of the excess MSE only
the reduced-rank parameter vector ${\bf w}_{k}$ because the MSE
attained upon convergence by the recursions, that work in
parallel, for ${\bf w}_{k}$ and ${\bf v}_{k}$ should be the same.
Here, we will drop the user $k$ index for ease of presentation.
Consider the MSE at time $i+1$ as:
\begin{equation}
\epsilon(i+1) = E[||b(i+1) - {\bf w}^{H}(i+1)\bar{\bf
r}(i+1)||^{2}]
\end{equation}
By using ${\bf w}(i+1)={\bf w}_{opt} + {\bf e}_{w}(i+1)$ and the
independence assumption the MSE becomes
\begin{equation}
\begin{split}
\epsilon(i+1)= & \epsilon_{min} - E[b(i+1)\bar{\bf
r}^{H}(i+1)]{\bf e}_{w}(i+1) \\ & - {\bf
e}_{w}^{H}(i+1)E[b^{*}(i+1)\bar{\bf r}(i+1)] \\ & + {\bf
w}_{opt}^{H} \bar{\bf R}(i+1) {\bf e}_{w}(i+1) \\ & + {\bf
e}_{w}^{H}(i+1) \bar{\bf R}(i+1){\bf w}_{opt} + \xi_{exc}(i+1)
\end{split}
\end{equation}
where $\epsilon_{min} = \sigma_{b} - E[b(i+1)\bar{\bf
r}^{H}(i+1)]{\bf w}_{opt} - {\bf w}_{opt}^{H} E[b^{*}(i+1)\bar{\bf
r}(i+1)] + {\bf w}_{opt}^{H}\bar{\bf R}(i+1){\bf w}_{opt}$ is the
MSE with the optimal reduced-rank receiver ${\bf w}_{opt}$ and the
optimal interpolator ${\bf v}_{opt}$ and $\xi_{exc}(i+1) = E[{\bf
e}_{w}^{H}(i+1) \bar{\bf R}(i+1) {\bf e}_{w}(i+1)]$ is the excess
MSE at time $i+1$. Since $\lim_{i \rightarrow \infty} E[{\bf
e}_{w}(i)]={\bf 0}$, we have
\begin{equation}
\lim_{i \rightarrow \infty} \epsilon(i+1) = \epsilon_{min} +
\lim_{i \rightarrow \infty} \xi_{exc}(i+1)
\end{equation}
Note that the second term in (67) is the steady-state excess MSE
due to adaptation, denoted by $\bar{\xi}_{exc}$ and which is
related to ${\bf w}$ by
\begin{equation}
\bar{\xi}_{exc}(\infty) = \lim_{i \rightarrow \infty} tr
E[\bar{\bf R} {\bf e}_{w}(i+1){\bf e}_{w}^{H}(i+1)]
\end{equation}
Let us define ${\bf R}_{e}(i)=E[{\bf e}_{w}(i){\bf
e}_{w}^{H}(i)]$, ${\bf R}_{e}=\lim_{i \rightarrow \infty} {\bf
R}_{e}(i)$ and use the property of trace to obtain
\begin{equation}
\bar{\xi}_{exc}(\infty) = tr E[\bar{\bf R} {\bf
R}_{e}]=vec^{H}(\bar{\bf R})vec({\bf R}_{e})
\end{equation}
At this point it can be noted that to assess
$\bar{\xi}_{exc}(\infty)$ it is sufficient to study ${\bf R}_{e}$,
that depends on the trajectory of the tap error vector. For
simplicity and similarly to \cite{xu&tsatsanis} we assume that
${\bf e}_{g}(i) \approx {\bf C}^{H}{\bf D}^{H} {\bf e}_{w}(i)$,
which is valid as the adaptation approaches steady state. Using
the expression of ${\bf e}_{w}(i+1)$, and taking expectation on
both sides of ${\bf e}_{w}(i+1){\bf e}_{w}^{H}(i+1)$, the
resulting matrix ${\bf R}_{e}(i+1)$ becomes
\begin{equation}
\begin{split}
{\bf R}_{e}(i+1) & \approx {\bf R}_{e}(i) - \mu ({\bf R}_{e}(i)
\bar{\bf R}(i)\boldsymbol{\Pi} + \boldsymbol{\Pi}\bar{\bf R} {\bf
R}_{e}(i)) \\ & \quad - \mu E[\boldsymbol{\Pi} {\bf e}_{w}(i){\bf
w}_{opt}^{H}\bar{\bf r}(i)\bar{\bf r}^{H}(i) \boldsymbol{\Pi} \\ &
\quad + \boldsymbol{\Pi} \bar{\bf r}(i)\bar{\bf r}^{H}(i){\bf
w}_{opt}{\bf e}_{w}^{H}(i) \boldsymbol{\Pi}] \\ &\quad + \mu^{2}
E[ \boldsymbol{\Pi} \bar{\bf r}(i)\bar{\bf r}^{H}(i)({\bf
w}_{opt}{\bf w}_{opt}^{H} \\ & \quad + {\bf R}_{e}(i)) \bar{\bf
r}(i)\bar{\bf r}^{H}(i)\boldsymbol{\Pi}]
\end{split}
\end{equation}
where $\boldsymbol{\Pi} = {\bf I} -{\bf DC}({\bf C}^{H} {\bf
D}^{H}{\bf D}{\bf C})^{-1}{\bf C}^{H}{\bf D}^{H}$. Since $\lim_{i
\rightarrow \infty} {\bf R}_{e}(i+1)= {\bf R}_{e}$ and $\lim_{i
\rightarrow \infty} E[{\bf e}_{w}(i)={\bf 0}$, taking limits on
both sides of (70) yields
\begin{equation}
\begin{split}
{\bf R}_{e} \bar{\bf R} \boldsymbol{\Pi} + \boldsymbol{\Pi}
\bar{\bf R}{\bf R}_{e} & \approx  \mu E[ \boldsymbol{\Pi} \bar{\bf
r}(i)\bar{\bf r}^{H}(i)({\bf w}_{opt}{\bf w}_{opt}^{H} \\ & \quad
+ {\bf R}_{e}(i)) \bar{\bf r}(i)\bar{\bf
r}^{H}(i)\boldsymbol{\Pi}]
\end{split}
\end{equation}
Here an expression for $\bar{\xi}_{exc}(\infty)$ can be obtained
by using the properties of the Kronecker product and arranging all
elements of a matrix into a vector column-wise through the $vec$
operation. Hence, the expression for the steady-state excess MSE
becomes:
\begin{equation}
\bar{\xi}_{exc}(\infty) = tr\Big[\bar{\bf R}{\bf R}_{e}\Big]= \mu
~ vec^{H}(\bar{\bf R}){\bf T}^{-1}{\bf a}
\end{equation}
where ${\bf T} = (\bar{\bf R}\boldsymbol{\Pi})^{T}\otimes{\bf I}
+{\bf I}\otimes(\boldsymbol{\Pi}\bar{\bf R}) - \mu
[\boldsymbol{\Pi}^{T} \otimes \boldsymbol{\Pi}] E\Big[({\bf
r}(i){\bf r}(i)^{H})^{T} \otimes ({\bf r}(i){\bf r}(i)^{H})\Big]$,
${\bf a} = \Big[(\boldsymbol{\Pi})^{T} \otimes
\boldsymbol{\Pi}\Big] E\Big[({\bf r}(i){\bf r}^{H}(i))^{T}
\otimes({\bf r}(i){\bf r}^{H}(i)) \Big]~vec({\bf w}_{opt}{\bf
w}_{opt}^{H})$ and $\otimes$ accounts for the Kronecker product.
The expression in (72) can be used to predict semi-analytically
the excess MSE, where the matrices $\bar{\bf R}$, ${\bf T}$ and
the vector ${\bf a}$ are computed through simulations.

\subsection{Transient Analysis and Convergence Speed}

With regard to convergence speed, adaptive receivers/filters have
a performance which is proportional to the number of adaptive
elements $M$ \cite{1,miller,haykin}. Assuming stationary noise and
interference, full-rank schemes with RLS algorithms take $2M$
iterations to converge, while SG algorithms require at least an
order of magnitude more iterations than RLS techniques
\cite{haykin}. In addition, it is expected that RLS methods do not
show excess MSE (when $\alpha=1$ and operating in a stationary
environment) and its convergence is independent of the eigenvalues
of the input correlation matrix.

With the proposed INT reduced-rank scheme, the convergence can be
made faster due to the reduced-number of filter coefficients and
the decimation factor $L$ can be varied in order to control the
speed and ability of the filter to track changing environments. In
Appendices II and III, we mathematically explain how the INT
structure can obtain gains in convergence speed over full-rank
schemes with SG and RLS algorithms, respectively.

For SG algorithms, the analysis of the transient components in
Appendix II of the INT scheme reveals that the speed of
convergence depends on the eigenvalue spread of the reduced-rank
covariance matrix. In principle, we cannot mathematically
guarantee that the INT always converges faster than the full-rank
but several studies that examine the eigenvalue spread of the
full-rank and the INT covariance matrix show that, for the same
data, the INT structure is able to consistently reduce the
eigenvalue spread found in the original data covariance matrix,
thus explaining its faster convergence in all analyzed scenarios.

For RLS techniques, the analysis of the transient components in
Appendix III guarantees mathematically that the INT is able to
converge faster due to the reduced number of filter elements and
we show that the INT with the RLS converges in about $2M/L$
iterations as compared to the full-rank that requires $2M$
iterations.

\section{Simulations}

In this section we investigate the effectiveness of the proposed
linear receiver structure and algorithms via simulations and
verify the validity of the convergence analysis undertaken for
predicting the MSE obtained by the adaptive algorithms. We have
conducted experiments under stationary and non-stationary
scenarios to assess the convergence performance in terms of
signal-to-interference-plus-noise ratio (SINR) of the proposed
structure and algorithms and compared them with other recently
reported techniques, namely adaptive versions of the MMSE
\cite{miller} and CMV \cite{xu&tsatsanis} full-rank methods, the
eigen-decomposition (PC) \cite{bar-ness,wang&poor} , the partial
despreading (PD) \cite{singh&milstein} and the multi-stage Wiener
filter (MWF) \cite{goldstein} reduced-rank techniques with rank
$D$. Moreover, the bit error rate (BER) performance of the
receivers employing the analyzed techniques is assessed for
different loads, processing gains ($N$), channel paths ($L_{p}$)
and profiles, and fading rates. The DS-CDMA system employs Gold
sequences of length $N=31$ and $N=63$.

Because we focus on the downlink users experiment the same channel
conditions. All channels assume that $L_{p}=6$ as an upper bound
(even though the effective number of paths will be indicated in
the experiments). For fading channels, the channel coefficients
$h_{l}(i)=p_{l}\alpha_{l}(i)$ ($l=0,1,2$), where
$\sum_{l=1}^{L_{p}} p_{l}^{2}=1$ and $\alpha_{l}(i)$, is a complex
unit variance Gaussian random sequence obtained by passing complex
white Gaussian noise through a filter with approximate transfer
function $c/\sqrt{1-(f/f_{d})^{2}}$ where $c$ is a normalization
constant, $f_{d}=v/\lambda_{c}$ is the maximum Doppler shift,
$\lambda_{c}$ is the wavelength of the carrier frequency, and $v$
is the speed of the mobile \cite{8}. This procedure corresponds to
the generation of independent sequences of correlated unit power
Rayleigh random variables ($E[|\alpha ^2_{l}(i)| ]=1$). The phase
ambiguity derived from the blind channel estimation method in
\cite{power2} is eliminated in our simulations by using the phase
of { ${\bf g}(0)$} as a reference to remove the ambiguity and for
fading channels we assume ideal phase tracking and express the
results in terms of the normalized Doppler frequency $f_{d}T$
(cycles/symbol). Alternatively, differential modulation can be
used to account for the phase rotation. For the proposed
interpolated receivers structures we employ $M=(N+L_{p}-1)/L$
adaptive elements for $L=2,3,4,8$, and when $M$ is not an integer
we will approximate it to the nearest integer. For the full-rank
receiver we have $M=(N+L_{p}-1)$.

In the following experiments, it is indicated the type of adaptive
algorithms used and their mode of operation, i.e. training mode,
decision-directed mode and blind mode. For the training-based
algorithms, the receiver employs training sequences with $N_{tr}$
symbols and then switch to decision-directed mode. The full-rank
receiver is considered with the NLMS and RLS tecnhiques, the
interpolated receivers are denoted INT, the PC method
\cite{bar-ness} requires an SVD on the full-rank covariance matrix
and the subspace dimension is chosen as $D=K$. For the PD
approach, the columns of the projection matrix are non-overlapping
segments of ${\bf s}_{k}$, as described in \cite{singh&milstein},
whereas for the MWF and its SG and recursive adaptive versions
(MWF-SG and MWF-rec) \cite{goldstein} the number of stages $D$ is
optimized for each scenario. The RAKE receiver in supervised mode
uses the NLMS and the RLS techniques and the training sequence in
order to estimate its parameters. With respect to blind algorithms
and the full-rank receiver, the SG algorithm corresponds to the
one in \cite{xu&tsatsanis} with a normalized step size similar to
the one introduced in Section IV.A and the RLS corresponds to the
one reported in \cite{xu&tsatsanis}. The proposed interpolated
receiver, i.e. the INT, uses the CMV-SG and CMV-RLS algorithms
especially designed for it. The different receiver techniques,
algorithms, processing gain $N$, the decimation factor $L$ and
other parameters are depicted on the legends. The
eigen-decomposition based receiver of Wang and Poor
\cite{wang&poor} is denoted Subspace-W $\&$ P and employs an SVD
to compute its eigenvectors and eigenvalues. With regard to blind
channel estimation we employ the method in \cite{power2} for all
SG based receivers, while for the RLS based receivers we adopt
\cite{power2}. The blind MWF and its adaptive versions (blind
MWF-SG and blind MWF-rec) \cite{goldstein} have their rank $D$
optimized for each situation and employ the blind channel
estimation in \cite{power2} to obtain the effective signature
sequence in multipath. For the RAKE receiver \cite{8}, we also
employ the SG blind channel estimation of \cite{power2} when
compared to other SG based multiuser receivers, whereas for the
comparison with RLS based receivers we use its RLS version
\cite{power2}.

\subsection{MSE Convergence Performance: Analytical Results}

Here, we verify that the results (64) and (72) of the section on
convergence analysis of the mechanisms can provide a means of
estimating the excess MSE. The steady state MSE between the
desired and the estimated symbol obtained through simulation is
compared with the steady-state MSE computed via the expressions
derived in Section VI. In order to illustrate the usefulness of
our analysis we have carried out some experiments. The
interpolator filters were designed with $N_{I}=3$ elements and the
channels have $3$ paths with gains $0$, $-6$ e $-10$ dB,
respectively, where in each run the delay of the second path
($\tau_{2}$) is given by a discrete uniform random variable (r.
v.) between $1$ and $4$ chips and the third path is computed with
a discrete uniform r. v. between  $1$ and $(5-\tau_{2})$ chips in
a scenario with perfect power control.

\begin{figure}[!htb]
\begin{center}
\def\epsfsize#1#2{1\columnwidth}
\epsfbox{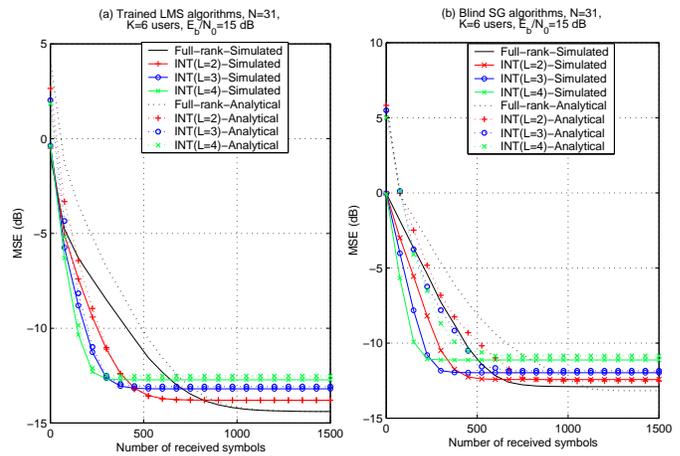} \caption{\small MSE convergence for analytical
and simulated results versus number of received symbols using (a)
trained LMS algorithms and (b) blind SG algorithms.}
\end{center}
\end{figure}

In the first experiment, we have considered the LMS algorithm in
trained mode and tuned the parameters of the mechanisms, in order
to achieve a low steady-state MSE upon convergence. The parameters
of convergence, i.e. $\mu$, are $0.05$, $0.06$, $0.075$ and $0.09$
for the full-rank and the INT with L=2,3 and 4, respectively, and
$\eta=0.005$ for the interpolator with all $L$. The results are
shown in Fig. 4 (a), and indicate that the analytical results
closely match those obtained through simulation upon convergence,
verifying the validity of our analysis.

In the second experiment, we have considered the blind SG
algorithm and tuned the parameters of the mechanisms, in order to
achieve a low steady-state MSE upon convergence, similarly to the
LMS case. The chosen values for $\mu$ are $0.0009$, $0.001$,
$0.0025$ and $0.004$ for the full-rank and the INT with L=2,3 and
4, respectively, and $\eta=0.005$ for the interpolator with all
$L$. The curves, depicted in Fig. 4 (b), reveal that a discrepancy
is verified in the beginning of the convergence process, when the
estimated covariance matrix is constructed with few samples. Also,
this mismatch between the theoretical and simulated curves is
explained by the fact that blind algorithms are more noisy than
trained techniques \cite{honig}. However, as time goes by and the
data record is augmented, the statistics of the signals is
acquired and the modelled and simulated MSE curves come to a
greater agreement.

\subsection{SINR Convergence Performance}

The SINR at the receiver end is used here to assess the
convergence performance of the analysed methods. In the following
experiments we will assess the SINR performance of the analyzed
adaptive receiver techniques and their corresponding algorithms,
namely, the proposed interpolated receiver, the PC, the PD, the
MWF and the RAKE. We remark that the parameters of the algorithms
have been tuned in order to optimize performance and the receiver
parameters have been carefully chosen to provide a fair comparison
amongst the analyzed methods.

\begin{figure}[!htb]
\begin{center}
\def\epsfsize#1#2{1\columnwidth}
\epsfbox{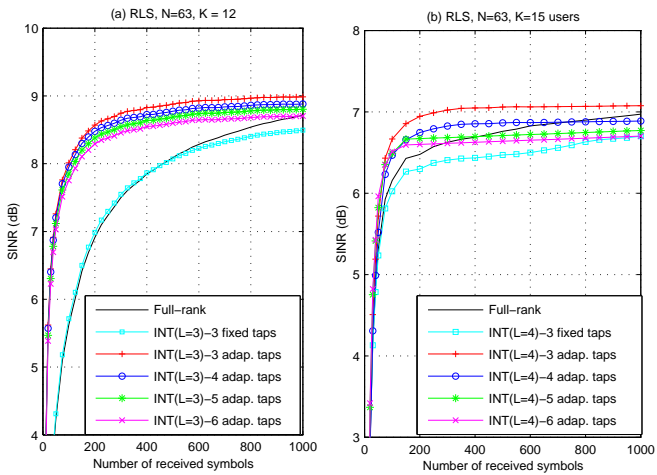} \caption{\small Design of interpolator filters
to obtain the best dimensions for $N_{I}$ with random $3$-path
channel parameters (r. v. between $-1$ and $1$) as given in
Section VI.A and the scenario has equal power users. (a) trained
RLS type algorithms at $E_{b}/N_{0}=12$ dB (b) blind CMV-RLS type
algorithms at $E_{b}/N_{0}=15$ dB.}
\end{center}
\end{figure}

Firstly, let us consider the issue of how long should be the
interpolator filter. Indeed, the design of the interpolator filter
is a fundamental issue in our approach because it affects its
convergence and BER performance. In order to obtain the most
adequate dimension for the interpolator filter ${\bf v}_{k}$, we
conducted experiments with values ranging from $N_{I}=3$ to
$N_{I}=6$, as the ones shown in Fig. 5 for the supervised and
blind modes with the RLS, respectively. The results indicate that
SINR performance was not sensitive to an increase in the number of
taps in ${\bf v}_{k}$ and the best results for all algorithms were
obtained with $N_{I}=3$. For this reason and to keep the
complexity low we selected $N_{I}=3$ for the remaining
experiments. We also remark that the simulation aided design of
the interpolator dimension was carried out for systems with
different $N$, $K$, $L$, channel profiles and fading rates,
indicating that $N_{I}=3$ is a satisfactory dimension. The SINR
convergence curves show that the proposed structure with adaptive
interpolators is considerably superior to the fixed interpolator
approach and to the full-rank receiver.

\begin{figure}[t]
\begin{center}
\def\epsfsize#1#2{1\columnwidth}
\epsfbox{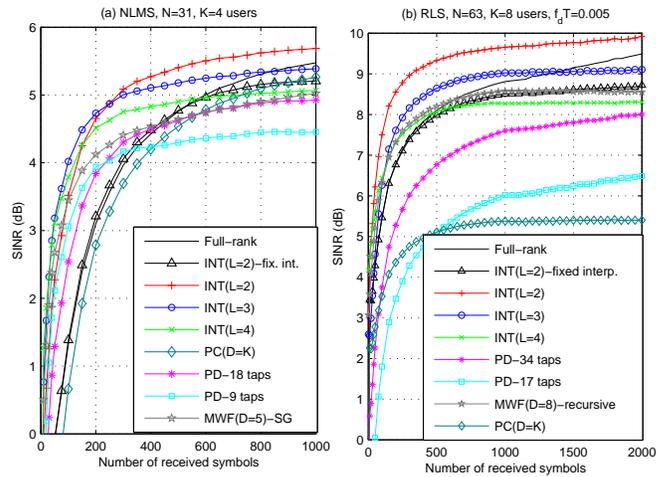} \caption{\small SINR performance of the
receivers with (a) NLMS, $E_{b}/N_{0}=8$ dB and three interferers
with power levels $7$ dB above the desired user and channel
parameters $p_{0} = 1$, $p_{2} = 0.5$ and $p_{4} = 0.3$ (spaced by
2$T_{c}$) and (b) RLS, $E_{b}/N_{0}=12$ dB and three interferers
with power levels $10$ dB above the desired user with fading and
channel parameters $p_{0} = 1$, $p_{2} = 0.7$ and $p_{4} = 0.5$
(spaced by 2$T_{c}$).}
\end{center}
\end{figure}

Fig. 6 illustrates experiments where the INT is compared to other
reduced-rank techniques in training and decision-directed modes.
In both experiments, a training sequence is provided to the
receivers with $200$ symbols, and then the algorithms switch to
decision-directed mode. The parameters of the receivers for all
methods were optimized and the results show that the proposed
structure with adaptive interpolators and $L=2$ achieves the best
performance and is significantly superior to the INT with a fixed
interpolator. The convergence performance of the INT for various
$L$ is superior to the full-rank one and to the PC and PD methods.
The PC method performs well when $K$ is small but it is
outperformed, both in terms of convergence speed and final SINR,
by the INT with $L=2,3$. The INT with $L=3$ and $L=4$ are also
superior to the PD method with $18$ and $9$ elements, whereas the
INT with $L=4$ has a performance comparable with the MWF adaptive
versions.

\begin{figure}[t]
\begin{center}
\def\epsfsize#1#2{1\columnwidth}
\epsfbox{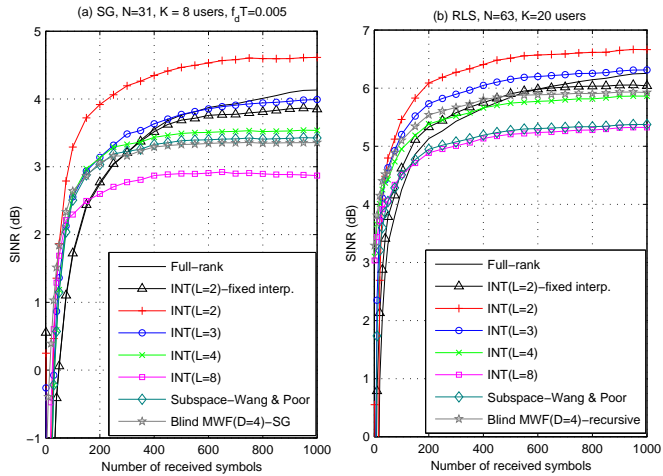} \caption{\small SINR performance of blind (a)
SG algorithms with channel parameters $p_{0} = 1$, $p_{2} = 0.5$
and $p_{4} = 0.5$ (spaced by 2$T_{c}$) where $2$ interferers work
at a power level $7$ dB above the desired user that operates at
$E_{b}/N_{0}=15$ dB with fading (b) RLS algorithm with
$E_{b}/N_{0}=15$ dB without fading, the $3$-path channel
parameters are random as in Section VI.A and the received powers
of the interferers are log-normal r. v. with associated standard
deviation $3$ dB.}
\end{center}
\end{figure}

In Fig. 7 the SINR performance of the analyzed receivers is
examined in blind mode. The parameters of the receivers for all
methods were optimized and the results show that the proposed
structure with adaptive interpolators and $L=2$ achieves the best
performance. The convergence performance of the novel structure
for various $L$ is superior to the full-rank one and to the other
methods. Note that subspace approach of Wang and Poor performs
very well for small $K$ but when $K$ is larger its performance
degrades considerably. The INT shows very good performance in all
situations and requires lower computational costs than the other
techniques.

\subsection{BER Performance}

In this section, the BER performance of the different receiver
techniques is investigated.  In Fig. 8, the BER curves for the RLS
algorithms in trained and decision-directed modes are shown. The
channel parameters are $p_{0} = 1$, $p_{1} = 0.7$ and $p_{2} =
0.5$, where in each run the delay of the second path ($\tau_{2}$)
is given by a discrete uniform r. v. between $1$ and $4$ chips and
the third path is computed with a discrete uniform r. v. between
$1$ and $(5-\tau_{2})$ chips. In these experiments the received
powers of the interferers are log-normal r. v. with associated
standard deviation $3$ dB. We remark that the proposed methods
also perform well with other channel profiles and fading rates.
The receivers are trained with $200$ symbols, are then switched to
decision-directed mode and process $2000$ symbols, averaged over
$200$ experiments with optimized parameters for each scenario. The
results show that the INT with $L=2$ achieves the best
performance, followed by the full-rank receiver, the INT with
$L=3$, the MWF, the PD approach, the INT with $L=4$, the PC and
the RAKE receiver.

\begin{figure}[!htb]
\begin{center}
\def\epsfsize#1#2{1\columnwidth}
\epsfbox{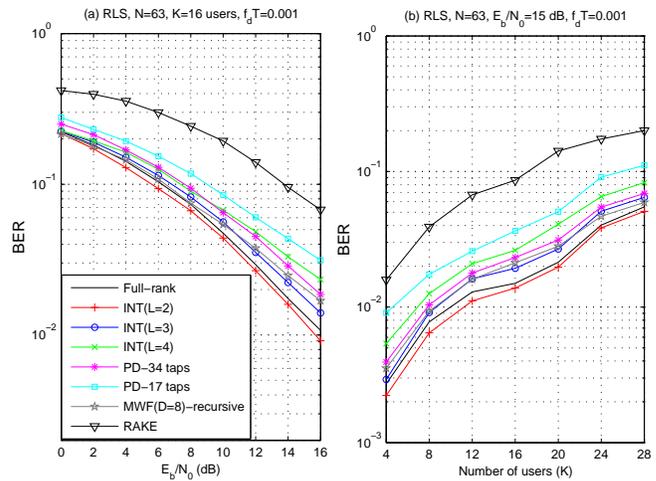} \caption{\small BER performance of trained RLS
algorithms versus (a) $E_{b}/N_{0}$ (b) number of users.}
\end{center}
\end{figure}

In Fig. 9, the BER curves for the RLS-type algorithms in blind
mode, respectively, are shown. The receivers process $2000$
symbols, averaged over $200$ experiments with optimized parameters
for each scenario. In these simulations the received powers of the
interferers are log-normal r. v. with associated standard
deviation $3$ dB. The results show that the INT with $L=2$
achieves the best performance, followed by the full-rank receiver,
the INT with $L=3$, the MWF, the INT with $L=4$, the subspace
receiver of Wang and Poor and the RAKE receiver. Note that the
receivers can accommodate more users and cope with larger systems
when working with RLS type algorithms and that the INT structure
with $L=4$ outperforms the RAKE and Wang and Poor's (for $K\geq8$)
receivers, the INT with $L=2$ outperforms the full-rank receiver
and the INT with $L=3$ has a very close performance to the
full-rank. The blind MWF versions are slightly inferior to the INT
with $L=3$ and suffer from the fact that tri-diagonalization does
not occur, deteriorating its performance.

\begin{figure}[!htb]
\begin{center}
\def\epsfsize#1#2{1\columnwidth}
\epsfbox{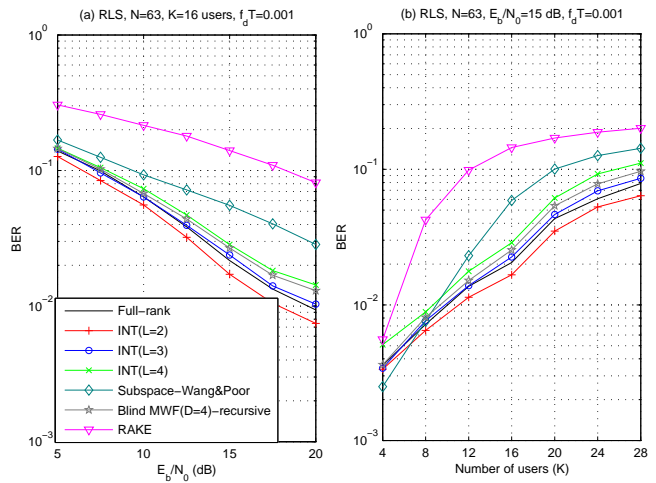} \caption{\small BER performance of blind RLS
algorithms versus (a) $E_{b}/N_{0}$ (b) number of users.}
\end{center}
\end{figure}

\section{Conclusions}

We proposed adaptive reduce-rank receivers for DS-CDMA based on
interpolated FIR filters with adaptive interpolators. The novel
receiver structure and algorithms were assessed in various
scenarios, outperforming previously reported techniques with a
very attractive trade-off between performance and complexity. An
analysis of the convergence properties of the method was
undertaken, indicating that the novel cost function does not
exhibit local minima. Furthermore, a convergence analysis of the
algorithms was shown to be valid in predicting the excess MSE upon
convergence for the blind and trained SG algorithms. In terms of
computational complexity, the AIFIR receivers are simpler than the
full-rank receiver, much simpler than reduced-rank
eigen-decomposition techniques and competes favorably with the
MWF. The BER performance of the interpolated receivers is superior
to the subspace receiver, the MWF and close to the full-rank one
even with $L=4$. Finally, with respect to convergence, the
proposed receivers exhibit a faster response and greater
flexibility than other analyzed methods since the designer can
choose the decimation factor depending on the need for faster
convergence or higher steady-state performance.

\begin{appendix}

\section{Proof of Lemma in Section IV.D}
Let ${\bf R}$ be a positive semi-definite Hermitian symmetric
matrix and its eigenvalues be ordered as $\lambda_{max} =
\lambda_{1}
> \lambda_{2} \geq \ldots \geq \lambda_{N-1} >
\lambda_{N}=\lambda_{min} \geq 0$ with corresponding eigenvectors
${\bf q}_{m}$ ($m=1,2,\ldots, N$). Consider an initial vector
$\hat{\bf v}(0)=\sum_{m=1}^{N} c_{m}{\bf q}_{m}$, where $c_{m}$
are scalars with $c_{1}=\hat{\bf v}^{H}(0){\bf q}_{1} \neq 0$.
Using the power iterations we have
\begin{equation}
\begin{split}
\hat{\bf v}(i)={\bf R}\hat{\bf v}(i-1)={\bf R}^{i}\hat{\bf v}(0) &
= c_{1}\lambda_{1}^{i}{\bf q}_{1} + c_{2}\lambda_{2}^{i}{\bf
q}_{2} + \ldots +
c_{N}\lambda_{N}^{i}{\bf q}_{N} \\
& = c_{1}\lambda_{1}^{i}({\bf q}_{1} +
c_{2}/c_{1}(\lambda_{2}/\lambda_{1})^{k}{\bf q}_{2} + \ldots +
c_{N}/c_{1}(\lambda_{N}/\lambda_{1})^{i}{\bf q}_{N}
\end{split}
\end{equation}
If we normalize the above equation we obtain
\begin{equation}
\frac{\hat{\bf v}(i)}{||\hat{\bf v}(i)||} = \frac{({\bf q}_{1} +
c_{2}/c_{1}(\lambda_{2}/\lambda_{1})^{k}{\bf q}_{2} + \ldots +
c_{N}/c_{1}(\lambda_{N}/\lambda_{1})^{i}{\bf q}_{N}}{||({\bf
q}_{1} + c_{2}/c_{1}(\lambda_{2}/\lambda_{1})^{k}{\bf q}_{2} +
\ldots + c_{N}/c_{1}(\lambda_{N}/\lambda_{1})^{i}{\bf q}_{N}||}
\end{equation}
and since $0 \leq \frac{\lambda_{m}}{\lambda_{1}} < 1$, for $2
\leq m \leq N$ it results that $\lim_{i \rightarrow \infty}
(\frac{\lambda_{m}}{\lambda_{1}})^{i} = 0$, for $2 \leq m \leq N$.
Thus, we conclude that
\begin{equation}
\lim_{i \rightarrow \infty} \frac{\hat{\bf v}(i)}{||\hat{\bf
v}(i)||} = {\bf q}_{max} = {\bf q}_{1}
\end{equation}
Now, let us make ${\bf A}={\bf I}-\nu{\bf R}$, where $\nu
=1/tr[{\bf R}]$ and whose eigenvalues are $\lambda_{m}' = 1 -
\frac{\lambda_{m}}{tr[{\bf R}]}$, $m=1,\ldots, N$. Since $tr[{\bf
R}] = \sum_{m=1}^{M} \lambda_{m} \geq \lambda_{1}=\lambda_{max}$,
it results that $0 \leq \lambda_{1}' \leq \lambda_{2}' \leq \ldots
\leq \lambda_{N-1}' < \lambda_{N}'$. Therefore, from the
development in (73)-(75), we have that the recursion $\hat{\bf
v}(i) = ({\bf I}-\nu\hat{\bf R})\hat{\bf v}(i-1)$, $i=1,2, \ldots$
results in
\begin{equation}
\lim_{i \rightarrow \infty} \frac{\hat{\bf v}(i)}{||\hat{\bf
v}(i)||} = {\bf q}_{N}' 
\end{equation}
where ${\bf q}_{N}'$ is the normalized eigenvector of ${\bf A}$
associated to $\lambda_{max}'=\lambda_{N}'=1 -
\frac{\lambda_{min}}{tr[{\bf R}_{\bf u_{k}}]}$, that is $({\bf
I}-\nu\hat{\bf R}){\bf q}_{N}'=(1 - \nu\lambda_{min}){\bf q}_{N}'$
and hence ${\bf q}_{N}'={\bf q}_{min}$.

\section{Convergence Speed of the INT Scheme with SG algorithms}

In this appendix we assess the convergence speed of the proposed
INT receiver scheme through the transient component analysis of SG
algorithms. By using a similar analysis to \cite{miller,haykin}
(see Chapter 9, pp. 390-404 and Appendix J, pp. 924-927 with the
solution to differential equations), let us express the excess MSE
in (59) as a function of its transient and steady-state
components:
\begin{equation}
\begin{split}
\xi_{exc}(i) & = \sum_{n=1}^{M/L}\lambda_{n}x_{n}(i) =
\boldsymbol{\lambda}^{H}{\bf x}(i) = \sum_{n=1}^{M/L}
\bar{c}_{n}^{~i} \boldsymbol{\lambda}^{H} {\bf g}_{n}{\bf
g}_{n}^{H} [{\bf x}(0) \\ & \quad- {\bf x}(\infty) ]  +
\xi_{exc}(\infty) = \xi_{trans}(i)+ \xi_{exc}(\infty)
\end{split}
\end{equation}
where $\bar{c}_{n}$ is the $n$th eigenvalue of an $M/L \times M/L$
matrix ${\bf T}$ whose entries are
\begin{equation}
t_{nj} = \left\{ \begin{array}{ll} (1 - \mu\lambda_{n})^{2} & n=j
\\ \mu^{2} \lambda_{n}\lambda_{j} & n \neq j
\end{array}\right.
\end{equation}
According to the above equation the speed of convergence of the
proposed INT structure for SG algorithms is given by the transient
component $\xi_{trans}(i) = \sum_{n=1}^{M/L} \bar{c}_{n}^{~i}
\boldsymbol{\lambda}^{H} {\bf g}_{n}{\bf g}_{n}^{H} [{\bf x}(0) -
{\bf x}(\infty) ]$ which can be alternatively expressed by:
\begin{equation}
\xi_{trans}(i) = \sum_{n=1}^{M/L} \gamma_{n} \bar{c}_{n}^{~i}
\end{equation}
where $\gamma_{n} = \boldsymbol{\lambda}^{H}{\bf g}_{n}{\bf
g}_{n}^{H} [ {\bf x}(0) - {\bf x}(\infty)]$. Note that the
transient component $\xi_{trans}(i) \rightarrow 0$ as $i
\rightarrow \infty$. By using the existing expression for the
transient component of the full-rank receiver described by
$\xi_{trans}^{full-rank}(i) = \sum_{n=1}^{M} \gamma_{n} c_{n}^{i}$
\cite{miller}, we can establish conditions for which the transient
component of the INT receiver defined in (79) can vanish faster,
i.e., the INT scheme converges faster. If the INT scheme reduces
the eigenvalue spread of its covariance matrix, we have for the
$i$th iteration
\begin{equation}
\sum_{n=1}^{M/L} \gamma_{n} \bar{c}_{n}^{~i} <  \sum_{n=1}^{M}
\gamma_{n} c_{n}^{i}
\end{equation}
The above condition states that the transient component of the
reduced-rank INT scheme has fewer decreasing modes and vanishes
before that of the full-rank structure. To verify (80), we studied
the eigenvalue spread of the covariance matrices of the INT and
the full-rank schemes in an extensive set of scenarios. In all
situations, the experiments indicate an increase in the
convergence speed and also that the INT can reduce the eigenvalue
spread of the full-rank scheme.

\section{Convergence Speed of the INT Scheme with RLS algorithms}

Here we evaluate the convergence speed of the proposed INT
receiver scheme through the MSE analysis of RLS algorithms. By
using a similar analysis to \cite{haykin} (see Chapter $13$, pp.
$573-579$) and replacing the expectation operator with time
averages, let us express weight error vector of the reduced-rank
INT least squares solution:
\begin{equation}
{\bf e}_{w}(i) = {\bf w}(i) - {\bf w}_{opt} = {\hat{\bar{\bf
R}}}^{-1}(i) \sum_{l=1}^{i} {\bf r}(l) e_{o}^{*}(l)
\end{equation}
Using the definition for the weight error correlation matrix ${\bf
K}(i)=E[{\bf e}_{w}(i){\bf e}_{w}^{H}(i)]$ \cite{haykin} we have:
\begin{equation}
{\bf K}(i) = E\Big[{\hat{\bar{\bf R}}}^{-1}(i) \sum_{l=1}^{i}
\sum_{j=1}^{i} {\bf r}(l)e_{o}^{*}(l)e_{o}(j){\bf r}^{H}(j)
{\hat{\bar{\bf R}}}^{-1}(i)\Big]
\end{equation}
Assuming that $e_{o}(i)$ is taken from a zero mean Gaussian
process with variance $\sigma^{2}$, we have
$E[e_{o}(l)e_{o}^{*}(j)] = \left\{
\begin{array}{ll} \sigma^2, & l=j
\\ 0, & l \neq j
\end{array}\right.$ and
\begin{equation}
{\bf K}(i) = \sigma^2 E\Big[{\hat{\bar{\bf R}}}^{-1}(i)
\sum_{l=1}^{i} \sum_{j=1}^{i} {\bf r}(l){\bf r}^{H}(j)
{\hat{\bar{\bf R}}}^{-1}(i)\Big] = \sigma^2 E\Big[{\hat{\bar{\bf
R}}}^{-1}(i)\Big]
\end{equation}
By invoking the independence theory and using the fact that the
estimate of the covariance matrix given by ${\hat{\bar{\bf
R}}}^{-1}(i)$ is described by a complex Wishart distribution
\cite{haykin} (see Section 13.6), the expectation of the time
averaged estimate ${\hat{\bar{\bf R}}}^{-1}(i)$ is exactly
\begin{equation}
E[{\hat{\bar{\bf R}}}^{-1}(i)] = \frac{1} {i-M/L -1 } \bar{\bf
R}^{-1}, ~~~~~~i>M/L +1
\end{equation}
where $\bar{\bf R}^{-1}$ is the theoretical reduced-rank
covariance matrix and thus
\begin{equation} {\bf K}(i) = \frac{\sigma^2 \bar{\bf
R}^{-1}} {i-M/L -1 } , ~~~~~~i>M/L +1
\end{equation}
Using the expression that describes the excess MSE in (58) we get
\begin{equation}
\xi_{exc}(i) = tr\Big[ \bar{\bf R} {\bf K}(i)\Big] =
\frac{\sigma^2 M/L} {i-M/L -1 } , ~~~~~~i>M/L +1
\end{equation}
The above result shows that the learning curve of the RLS
algorithm with the proposed reduced-rank structure converges in
about $2M/L$ iterations, in contrast to the RLS with the full-rank
scheme, that requires about $2M$ iterations \cite{haykin}. This
means that the proposed scheme converges $L$ times faster than the
full-rank approach with RLS techniques. Another observation from
(86) is that as $i$ increases the excess MSE tends to zero (for
$\lambda = 1$) and it is independent from the eigenvalue spread of
${\hat{\bar{\bf R}}}^{-1}(i)$.

\end{appendix}

{ \linespread{0.9}

\end{document}